\documentclass[aps,prd,twocolumn,nofootinbib,preprintnumbers]{revtex4-2}

\usepackage{amsmath,amssymb,amsfonts,mathtools,bm,booktabs,slashed,xcolor}
\usepackage{hyperref}
\usepackage{mathrsfs}
\usepackage{graphicx}
\usepackage{xspace}
\hypersetup{colorlinks=true,linkcolor=blue,citecolor=blue,urlcolor=blue}
\usepackage{tikz}
\usetikzlibrary{shapes.geometric,decorations.pathmorphing,decorations.markings,arrows.meta,calc}

\newcommand{\cC}{\mathcal{C}}

\newcommand{\cF}{\mathcal{F}}

\newcommand{\cH}{\mathcal{H}}

\newcommand{\cK}{\mathcal{K}}

\newcommand{\cN}{\mathcal{N}}
\newcommand{\cO}{\mathcal{O}}

\newcommand{\cV}{\mathcal{V}}

\newcommand{\AdS}{\mathrm{AdS}}
\newcommand{\dd}{\mathrm{d}}
\newcommand{\ee}{\mathrm{e}}
\newcommand{\ii}{\mathrm{i}}

\newcommand{\diag}{\mathrm{diag}}
\newcommand{\ord}{\mathcal{O}}

\newcommand{\eps}{\varepsilon}

\newcommand{\bdry}{\mathrm{bdry}}
\newcommand{\DDVCS}{\textsc{DDVCS}\xspace}
\newcommand{\DVCS}{\textsc{DVCS}\xspace}
\newcommand{\TCS}{\textsc{TCS}\xspace}
\newcommand{\hH}{\widehat{\cH}}
\newcommand{\beps}{\bar\epsilon}
\newcommand{\wt}[1]{\widetilde{#1}}
\DeclareMathOperator{\arccosh}{\arccosh}

\begin{document}

\title{Holographic Open/Closed Exchange in Double Deeply Virtual Compton Scattering: Fixed--$j$ Structural Matching to the $\pm$-Basis Wilson Kernels}

\author{Kiminad A.~Mamo}
\affiliation{Department of Physics, University of Connecticut, Storrs, CT 06269-3046, USA}
\email{ska25005@uconn.edu}
\date{\today}

\begin{abstract}
We show that fixed--$j$ holographic double deeply virtual Compton scattering (DDVCS), and its DVCS limit, give a fixed-scale structural match to the singlet conformal-OPE Wilson-kernel family of QCD in the leading-twist unpolarized singlet vector channel. The open-string $C_1(\delta,\vartheta)$ hypergeometric kernel was derived by Nishio--Watari; the new closed-string result is that the BPST upper Witten vertex gives the same family with a different near-boundary power count, because $\Delta_c(j)$ is fixed by the BPST trajectory rather than by open-Regge data. The Mellin exponent is therefore derived as $\delta_c(j)=j+\Delta_c(j)-2=2j+\gamma_c(j)$ rather than inserted by hand. The even open branch gives the parallel unprotected counterpart in the projected singlet vector amplitude. At $Q=\mu=\mu_0=\mu_\ast$, the point is not merely that equal-endpoint evolution is trivial; it is that the ultraviolet Witten vertex has already produced the fixed--$j$ conformal kernel before the lower conformal moment is matched. In the conformal partial-wave/CS representation, beta-proportional conformal-anomaly terms and scheme transformations are coefficient/evolution bookkeeping away from the matching point, not replacements for the projected fixed-scale hypergeometric $\eta/\xi$ Wilson-kernel family. The protected/unprotected $j=2$ split identifies the protected closed branch with the $(-)$ conformal partial wave and the even open branch with the unprotected $(+)$ counterpart; these are mixed singlet eigenchannels, not literal unmixed quark/gluon operators at finite $N_c$. Thus, once the lower Witten vertex is matched to constrained conformal moments, the DVCS/DDVCS deconvolution problem is organized in a physical operator basis rather than in arbitrary $x$-space profiles.
\end{abstract}

\maketitle

\section{Introduction}
\label{sec:intro}

Deeply virtual Compton scattering (\DVCS), timelike Compton scattering (\TCS), and double deeply virtual Compton scattering (\DDVCS) are among the cleanest probes of generalized parton distributions (GPDs), which were introduced in the original momentum fraction $x$-space/nonforward-distribution formulations by M\"uller \textit{et al.}, by Ji and by Radyushkin, and subsequently developed into a broad QCD framework~\cite{Muller:1994ses,Ji:1996ek2,Ji:1996ek,Radyushkin:1996ru,Radyushkin:1996nd,Radyushkin:1997ki,Diehl:2003ny,Belitsky:2005qn}. At leading power, the \DVCS amplitude admits a collinear factorization theorem: the short-distance current-current correlation function is separated from the long-distance hadronic matrix element in the generalized Bjorken regime, to all orders in perturbation theory~\cite{Ji:1998xh,Collins:1998be}. The same amplitude can be organized not only in the original momentum-fraction language but also in the conformal-basis / conformal-moment language, where complex conformal spin diagonalizes the leading-order singlet evolution and makes the operator product expansion of the two currents particularly transparent~\cite{Mueller:1997hs,Belitsky:1997rh,Mueller:2005ed,Kirch:2005hu,Manashov:2005qm,Kumericki:2006xx,Kumericki:2007sa}. This operator-basis organization matters experimentally: in a JLab or EIC extraction the quantity to constrain is a physical conformal moment multiplying a fixed hard kernel, not an arbitrary $x$-space skewness profile. Recent two-loop work has made this conformal-basis description increasingly explicit, including vector, axial-vector, transversity, \DDVCS, and conformal-moment coefficient functions~\cite{Braun:2020yib,Braun:2021ysj,Braun:2022nnlo,Ji:2023transversity,Braun:2024ddvcs,Braun:2026dvcsmom}. Recent \DDVCS phenomenology and sensitivity studies make the same kinematic point experimentally: the second photon virtuality opens access to GPD information beyond the on-shell \DVCS/\TCS lines, with feasibility studied for Jefferson Lab/CEBAF, CLAS/SoLID, and EIC settings~\cite{Deja:2023ahc,Alvarado:2025prc,Alvarado:2026ggy,Accardi:2023chb}. String/holographic trajectories enter this experimental problem as a fixed--$j$ Witten-diagram mechanism for producing the conformal moments and kernels that the operator basis already asks for. The same framework has been used in the parametrization and global extraction of small-skewness GPDs~\cite{Guo:2022ump0,Guo:2023ump,Guo:2025muf}, and in all-skewness string/holographic conformal-moment GPD parametrizations~\cite{Mamo:2024jwp,Mamo:2024vjh,Hechenberger:2025rye,Hechenberger:2025wnz}.

The conformal/Gegenbauer basis is more than a mathematical trick or one possible orthogonal basis in which to expand GPDs or Compton form factors. In the measured Compton form factors the underlying GPDs enter through convolution integrals, so the inverse problem is a deconvolution problem: many different $x$-space GPD profiles can lead to similar \DVCS/\DDVCS amplitudes~\cite{Bertone:2021yyz}. The conformal basis is distinguished because its moments are matrix elements of local twist-two conformal operators. For the physical even-spin local moments, their skewness dependence is therefore tied to Lorentz covariance and polynomiality, and at leading order the singlet evolution is diagonal in the $\pm$ basis. Holographic QCD adds a second reason for using the same basis: the lower Witten vertex provides a holographic representation of the same fixed--$j$ conformal moments geometrically, with model-dependent input but with polynomiality constraints kept manifest. Once such moments are represented by a holographic-string construction, the deconvolution problem is recast as the determination of physical conformal-moment data multiplying a fixed Wilson kernel, not as the unconstrained reconstruction of arbitrary $x$-space profiles. This is precisely the logic behind all-skewness holographic GPD parametrizations~\cite{Mamo:2024jwp,Mamo:2024vjh,Hechenberger:2025rye,Hechenberger:2025wnz}. The result shown below explains why that strategy is not an arbitrary basis choice: the UV Witten vertex itself selects the same conformal Wilson-kernel family as the QCD OPE.

From the holographic side, a classic ultraviolet anchor is the vacuum vector-current two-point function. In bottom-up AdS/QCD, the on-shell bulk action reproduces the logarithmic $Q^2$ behavior of the QCD current-current correlator and fixes the bulk gauge coupling~\cite{Erlich:2005qh}. In the soft-wall model, the same current is sourced by the bulk-to-boundary propagator of Grigoryan and Radyushkin~\cite{Grigoryan:2007vg}. This vacuum matching fixes the non-normalizable current source; it does not by itself fix spin--$j$ source normalizations, trajectory data, or hadronic conformal moments. The claim below is that the same current insertion, embedded in the fixed--$j$ hadronic Witten diagram, yields the QCD conformal Wilson-kernel family, while the exchanged-operator normalization and lower conformal moment remain matched or modeled inputs. At high energy and small-$x$, the closed-string channel is anchored by the BPST graviton trajectory~\cite{Brower:2006ea}, while holographic \DVCS amplitudes have been analyzed in conformal Regge theory by Costa and Djuri\'c, in related work by Brower \textit{et al.}~\cite{Costa:2012cb,Brower:2012mk}, and on $s$-channel/large-$x$ scalar target in~\cite{Marquet:2010sf}. The connection between holographic \DVCS/GPD amplitudes and conformal-OPE language, including the open-string sector, was made especially transparent by Nishio and Watari~\cite{Nishio:2014rya,Nishio:2014eua}. Appendix~B, especially Eqs.~(285)--(289) of~\cite{Nishio:2014rya,Nishio:2014eua}, defines the $C_1(\delta,\vartheta)$ AdS integral and rewrites it as the same Gauss-hypergeometric conformal-OPE kernel used below. More broadly, graviton-like effective descriptions of \(2^{++}\) tensor mesons have long appeared in AdS/QCD~\cite{Katz:2005ir}, and tensor-meson exchange remains phenomenologically relevant in current hadronic analyses~\cite{Mager:2025pvz}.

The novelty relative to Nishio--Watari is therefore sharply delimited. Their Appendix~B supplies the open-string $C_1$ hypergeometric kernel. Here the same kernel is derived for the closed BPST exchange for the first time in this fixed--$j$ Compton setting, with the exponent $\delta_c(j)=j+\Delta_c(j)-2$ fixed by the explicit BPST upper-vertex $z$-power counting. This closed result is not an immediate corollary of the open one: the closed exponent follows from the BPST boundary mode and trajectory data and therefore requires an independent near-boundary derivation with no open-string input. The paper then places the closed BPST and even open Nishio--Watari branches in the modern QCD $\pm$ eigenbasis, keeps the finite forward normalizations $c_j^\pm$, interprets the lower Witten vertex as the nonperturbative conformal moment, and uses the protected/unprotected $j=2$ split to identify the protected branch and supply the analytic-continuation prescription for the structural closed/open $\leftrightarrow$ $(-)/(+)$ assignment. In one sentence: the novelty is not the open \(C_1\) identity itself, but the independent closed BPST fixed--\(j\) upper-Witten-vertex derivation of the same conformal Wilson-kernel family, with the Mellin label fixed by \(z\)-power counting, and the resulting structural open/closed \(\leftrightarrow(+)/(-)\) operator-basis mapping relevant for \DDVCS/\DVCS deconvolution. This is a structural matching result, not a rediscovery of the open-string hypergeometric identity.

The central question of this paper is whether the holographic description of \DDVCS and its \DVCS limit reproduces the same fixed--$j$ hard-kernel structure that appears in the conformal-basis factorization formula of QCD. The construction below derives the hard kernel from the ultraviolet Witten vertex and reserves infrared sensitivity for the lower conformal moment. A positive answer gives holographic QCD a precise role in the twist-two singlet Compton amplitude: it becomes a concrete realization of the fixed--$j$ factorized structure of the conformal OPE, with Regge behavior, vector-meson dominance, and large-$N_c$ organization tied to the same operator-basis language.

The cleanest route is to begin with the fixed--$j$ $t$--channel Witten diagram and factorize it before performing the Sommerfeld--Watson resummation. In that order of limits the physics is transparent. The upper vertex depends only on the two virtual photons and the near-boundary behavior of the exchanged spin--$j$ mode. The lower vertex is the nucleon matrix element of the same spin--$j$ operator. The spin projector supplies the kinematical polynomial in the skewness variables. What remains after this separation is the holographic analogue of a fixed--$j$ Compton form factor.

The crucial result is that the upper holographic impact factor is universal. In the conformal limit it is determined entirely by the pure-$\AdS$ photon wave functions and yields a closed-form Gauss hypergeometric function of $\eta^2/\xi^2$. The lower hadronic vertex carries all infrared model dependence and may be evaluated in the soft-wall model, or in any other infrared completion, without altering the hard kernel. This already gives a clean holographic factorization statement, see Figures~\ref{TWITTEN-2} and \ref{fig:witten-factorized-closed}. The stronger claim made here is that the same hypergeometric family is precisely the one that appears in the conformal-basis Wilson coefficients of the singlet vector Compton form factor in perturbative QCD. Evidence that the requisite near-boundary factorization of the bulk-to-bulk propagator is not an artifact of a particular infrared model can be traced through the pure-AdS / hard-wall analysis of Nishio and Watari, the coordinate-space graviton analysis of Hatta and Yang, and the hard-/soft-wall Witten-diagram treatment of heavy quarkonium photoproduction by Mamo and Zahed~\cite{Nishio:2014rya,Nishio:2014eua,Hatta:2018jpsi,Mamo:2019mka}.

In concrete terms, the UV part of the holographic Witten diagram produces the same fixed--$j$ conformal Wilson-kernel family that appears in the conformal-basis collinear-factorization formula, the IR part provides the corresponding matched nonperturbative conformal moment, and the first physical even moment $j=2$ anchors the closed/$(-)$ and open/$(+)$ channel assignment dynamically rather than conventionally. At \(Q=\mu=\mu_0=\mu_\ast\), the evolution factors are identities, but the fixed-scale coefficient is the hypergeometric Wilson kernel multiplied by the finite normalization \(c_j^\pm\), and the lower conformal moment remains a matched input. The scope of this statement---fixed $j$, fixed scale, TT projection, and leading twist in the collinear window---is stated explicitly in Sec.~\ref{subsec:scope-limitations}.

More precisely, in the collinear matching window
\begin{align}
&Q_1^2\sim Q_2^2\sim \widetilde Q^{\,2}\equiv Q^2,
\qquad
-t,M_N^2\ll Q_i^2,\nonumber\\
&\xi,\eta\ll1,
\qquad
\vartheta\equiv\frac{\eta}{\xi}=\mathcal O(1),
\qquad i=1,2.
\label{eq:intro-matching-window}
\end{align}
The projection used throughout is the transverse--transverse scalar component of the current correlator. In this collinear window the matching is applied after this TT scalar projection: $F_{z\mu}$-induced structures involving $K_0$ or $z\partial_zK_1$ project onto longitudinal or distinct invariant amplitudes and are not part of the scalar kernel matched below.
At the matching point
\begin{equation}
Q=\mu=\mu_0=\mu_\ast,
\label{eq:intro-matching-point}
\end{equation}
we will show that the even open-string channel matches the $+$ conformal-basis Wilson coefficient and conformal moment, while the closed-string channel matches the corresponding $-$ objects. The natural arena for stating this result is \DDVCS, where both photons are off shell and the conformal structure is fully visible; the \DVCS limit may then be taken only after the fixed--$j$ analysis. This ordering matters physically: when both photons remain off shell, the ratio $\eta^2/\xi^2$ appears in the hard kernel in its fully visible conformal form, whereas taking one photon on shell too early hides part of that structure even though it does not change the underlying fixed--$j$ operator statement~\cite{Ji:1996ek,Braun:2024ddvcs}.

The most decisive diagnostic is the first physical even moment, $j=2$. On the perturbative side, the $(-)$ singlet anomalous dimension vanishes at $j=2$ as the leading-order manifestation of the momentum-sum rule, while the $(+)$ eigenvalue remains finite; the explicit formulas are collected in Appendix~\ref{app:pqcd}, especially Eqs.~\eqref{eq:app-pqcd-gamma-pm-at2} and \eqref{eq:app-pqcd-gamma-pm-near2}. On the holographic side, the closed branch passes through the bulk graviton and therefore through the conserved total energy--momentum tensor, so the corresponding anomalous dimension vanishes at the same point. The open branch has no such symmetry protection at $j=2$; in the large-$\lambda$ regime its value is finite and positive, while any accidental zero at a special finite value of $\lambda$ would only produce a temporary degeneracy at that point, not a Ward identity. The open/$+$ and closed/$-$ assignment is therefore not a fit-driven analogy but a structural prescription anchored by symmetry and by the analytic behavior of the two branches.

It is equally important to state the role of conformality precisely. The conformal fixed-point formulas provide the cleanest analytic representation of the hard kernel because they display the scale dependence as a pure power law. No claim is made that real QCD is near a conformal fixed point; the conformal expressions are used as the conformal partial-wave/CS representation, where the hard kernel is most transparent. The matching itself is imposed at the single scale \eqref{eq:intro-matching-point}. At that point, the matched object is the fixed--$j$ conformal Wilson-kernel family. Beta-proportional conformal-anomaly terms, finite scheme transformations, and logarithmic running are part of the coefficient/evolution bookkeeping away from the matching point. They refine the anomalous dimensions, finite coefficient normalizations, and evolution factors, but they are not an alternative fixed-scale $\eta/\xi$ kernel for the projected invariant. Recent NNLO coefficient functions for \DVCS and \DDVCS sharpen the perturbative input for $\gamma_j^\pm(\alpha_s)$, $c_j^\pm(\alpha_s)$, and evolution effects~\cite{Braun:2020yib,Braun:2021ysj,Braun:2022nnlo,Ji:2023transversity,Braun:2024ddvcs,Braun:2026dvcsmom}. They refine the perturbative data entering the same conformal partial wave rather than replacing the fixed-scale hypergeometric kernel.

The paper is conceptual rather than numerical. It determines the operator meaning of the holographic fixed--$j$ amplitude, isolates the universal part from the infrared-sensitive part, and places the resulting nonperturbative moments in the same language used by collinear factorization and, ultimately, by lattice-QCD determinations of moments and form factors. In this sense it is a structural paper about what the holographic amplitude is computing.

The main body is organized around this narrative. Section~\ref{sec:main-statement} states the fixed--$j$ holographic amplitude and the precise matching dictionary to the conformal-basis pQCD result. Section~\ref{sec:holo-main} derives the universal holographic hard kernel from the Witten diagram and isolates the model-dependent lower conformal moments. Section~\ref{sec:pqcd-main} writes the singlet pQCD Compton form factor in the $\pm$ basis, formulates the fixed--$j$ matching as a single-scale statement, and explains why the fixed-point formulas are a representation choice rather than an additional dynamical assumption. Section~\ref{sec:physical-meaning} reconstructs the physical Compton form factor and discusses the operator interpretation, the \DDVCS$\to$\DVCS transition, and the phenomenological scope. Appendix~\ref{app:holo} contains the holographic proof in full detail, including the near-boundary factorization, the explicit open-channel vertex, and the open/closed replacement rule, while Appendix~\ref{app:pqcd} collects the complete pQCD derivation, including the diagonalization of singlet evolution, the normalization constants $c_j^\pm$ fixed from forward DIS, the running-coupling implementation, the quark--gluon basis, the large-$N_c$ limit, and the expansion near $j=2$.

\section{Central fixed--$j$ statement, scope, and matching dictionary}
\label{sec:main-statement}

We now state the object that is being matched before turning to its holographic and perturbative derivations. Throughout this section the object is the leading-twist unpolarized singlet vector invariant amplitude in the collinear window \eqref{eq:intro-matching-window}. Full Compton-tensor components, helicity structures, axial-vector or transversity channels, contact terms, target-mass effects, and higher-twist corrections lie outside this scalar projection and are summarized in Sec.~\ref{subsec:scope-limitations}. Within this restriction the upper current insertion is always the TT-projected scalar component. Terms in a fully gauge-invariant $VV\cO_j$ coupling that contain $F_{z\mu}$, $K_0$, or $z\partial_zK_1$ feed longitudinal or distinct tensor projections; they are not discarded from the full theory, only projected outside the invariant amplitude being matched. The schematic projected upper-vertex statement is
\begin{equation}
-\frac12 g^T_{\mu\nu}\,\cV^{\mu\nu}_{VV\cO_j}\Big|_{\rm lead}
=\cN_j\,\cV(Q_1,z)\cV(Q_2,z)
+\ord\!\left(\frac{M_N^2}{Q^2},\frac{-t}{Q^2}\right),
\label{eq:main-tt-upper-projection}
\end{equation}
while the complementary $F_{z\mu}$-induced terms define LL, TL, or helicity-distinct kernels. This is the same TT/LL separation used in the Nishio--Watari decomposition and in the holographic electroproduction projection of Sec.~II.D, especially Eq.~(II.17), of Ref.~\cite{Mamo:2021tzd}.

We use the average hadron and photon momenta
\begin{equation}
P\equiv \frac{p+p'}{2},
\qquad
\widetilde q\equiv \frac{q_1+q_2}{2},
\qquad
\Delta\equiv p-p'=q_2-q_1,
\end{equation}
with invariants
\begin{widetext}
\begin{equation}
p^2=p'^2=M_N^2,
\qquad
q_1^2=-Q_1^2,
\qquad
q_2^2=-Q_2^2,
\qquad
\widetilde q^{\,2}=-\widetilde Q^2,
\qquad
t\equiv \Delta^2<0,
\qquad
s\equiv (p+q_1)^2,
\end{equation}
\end{widetext}
and the standard off-forward variables
\begin{equation}
\eta\equiv \frac{\Delta\!\cdot\!\widetilde q}{2P\!\cdot\!\widetilde q},
\qquad
\xi\equiv \frac{\widetilde Q^2}{2P\!\cdot\!\widetilde q},
\qquad
\vartheta\equiv \frac{\eta}{\xi},
\qquad
\chi\equiv \frac{4M_N^2\eta^2}{-t}.
\end{equation}
For definiteness, introduce lightlike collinear vectors $p_c^\mu$ and $n_c^\mu$ with $p_c^2=n_c^2=0$ and $p_c\!\cdot n_c=1$, and define the transverse metric
\begin{equation}
g_T^{\mu\nu}=g^{\mu\nu}-p_c^\mu n_c^\nu-p_c^\nu n_c^\mu .
\label{eq:TTmetric}
\end{equation}
The scalar invariant matched in this paper is the TT projection of the singlet vector Compton tensor,
\begin{equation}
\widehat{\cH}\equiv -\frac12 g_{T\mu\nu}T_{\rm sing,V}^{\mu\nu}
+\ord\!\left(\frac{M_N^2}{Q^2},\frac{-t}{Q^2}\right),
\label{eq:TTscalar-invariant}
\end{equation}
with the omitted terms belonging to longitudinal, helicity-flip, target-mass, or higher-twist invariant amplitudes. All occurrences of the upper photon vertex below are understood after this projection.
Throughout, \(X=o,c\) labels the even open and closed channels, respectively. We also use the convention
\begin{equation}
\cK_j^{(\gamma)}(\vartheta)
\equiv
{}_2F_1\!\left(
\frac j2+\frac\gamma4,
\frac{j+1}{2}+\frac\gamma4;
j+\frac32+\frac\gamma2;
\vartheta^2
\right),
\label{eq:main-universal-kernel}
\end{equation}
and reserve the term ``Wilson kernel'' for this hypergeometric functional dependence. The corresponding ``Wilson coefficient'' is the full normalized coefficient, including \(c_j^\pm\), scale/evolution factors, and basis-normalization conventions.
\begin{table*}[t]
\centering
\caption{Fixed--$j$ matching dictionary. The table summarizes the structural assignments used below; the derivation and normalization conventions are given in the text and appendices.}
\begin{tabular}{p{0.21\textwidth}p{0.22\textwidth}p{0.25\textwidth}p{0.23\textwidth}}
\toprule
Holographic object & QCD conformal-basis object & Diagnostic & Structural role \\
\midrule
closed BPST branch & $(-)$ singlet eigenchannel & $\gamma_c(2)=0\leftrightarrow\gamma_2^{-,(0)}=0$ & protected energy--momentum branch \\
even Nishio--Watari open branch & $(+)$ singlet eigenchannel & no Ward-identity zero at $j=2$ & unprotected even open Reggeon branch \\
upper Witten vertex & conformal Wilson kernel & same $\cK_j^{(\gamma)}(\eta/\xi)$ & universal projected hard factor \\
lower Witten vertex & conformal moment & matched nonperturbative input & IR-sensitive operator matrix element \\
\bottomrule
\end{tabular}
\label{tab:fixedj-dictionary}
\end{table*}

In the twist-two unpolarized sector, the fixed--$j$ holographic amplitudes may be written in the compact channel-unified form
\begin{widetext}
\begin{equation}
\widehat{\cH}^{(X)}_{\rm holo}(j)
=
\xi^{-j}
\left(\frac{\mu}{Q}\right)^{\gamma_X(j)}
\cK_j^{(\gamma_X(j))}(\vartheta)
\left(\frac{\mu_0}{\mu}\right)^{\gamma_X(j)}
\phi_0^{X}(j,\widetilde\delta_X,\widetilde\delta_0)
\widehat d_j(\eta,t)
\mathfrak g_X
\widehat{\cF}_N^{(X)}(j;t),
\qquad X=o,c,
\label{eq:main-holo-fixedj}
\end{equation}
\end{widetext}
where
\begin{equation}
\mathfrak g_o\equiv 1,
\qquad
\mathfrak g_c\equiv \frac{\widetilde g_5^2}{g_5^2},
\label{eq:main-gX-def}
\end{equation}
and
\begin{equation}
\widehat d_j(\eta,t)
\equiv
{}_2F_1\!\left(
-\frac{j}{2},\frac{1-j}{2};\frac12-j;-
\frac{4M_N^2\eta^2}{-t}
\right)
\label{eq:main-dhat-def}
\end{equation}
is the kinematical polynomial produced by the spin--$j$ projector. For physical even $j$, it is the finite polynomial in $4M_N^2\eta^2/(-t)$ obtained by contracting the symmetric-traceless spin--$j$ projector between the two hadron momenta in the off-forward $t$ channel; Eq.~\eqref{eq:main-dhat-def} is the analytic continuation of that polynomial used in the fixed--$j$ Regge representation. It is part of the lower matrix-element kinematics and does not modify the upper $Q^2$ and $\eta/\xi$ Wilson kernel. Equation~\eqref{eq:main-holo-fixedj} is the precise fixed--$j$ object whose universal kernel is being matched. The quantity $\phi_0^X$ is the residual UV source normalization, while $\widehat{\cF}_N^{(X)}(j;t)$ is the bottom conformal moment supplied by the hadronic vertex.

The point of Eq.~\eqref{eq:main-holo-fixedj} is that the full dependence on $Q^2$ and $\eta/\xi$ factorizes into the universal hypergeometric hard kernel. All infrared model dependence is pushed into $\widehat{\cF}_N^{(X)}$. The holographic derivation in Section~\ref{sec:holo-main} and Appendix~\ref{app:holo} shows that the Mellin exponent is fixed to be
\begin{equation}
\delta_X(j)\equiv j+\Delta_X(j)-2=2j+\gamma_X(j),
\qquad X=o,c,
\label{eq:main-delta-def}
\end{equation}
where $\Delta_X(j)$ is the dimension of the exchanged spin--$j$ mode and $\gamma_X(j)$ its anomalous contribution.

To compare directly with the perturbative literature, we organize the singlet QCD amplitude in the $\pm$ basis of Ref.~\cite{Kumericki:2007sa}, especially Sec.~4.1, and make only one convention change:
\begin{equation}
j_{\rm ref}\to j-1,
\end{equation}
so that our label $j$ coincides with the physical spin-$j$ exchanged in the fixed-$j$ channel. Recent conformal-moment calculations for \DVCS and \DDVCS can be interfaced naturally with the same organization~\cite{Braun:2024ddvcs,Braun:2026dvcsmom}.

The fixed--$j$ singlet vector Compton form factor in the conformal basis that diagonalizes the LO singlet evolution takes the form
\begin{equation}
\widehat{\cH}^{\rm sing}_{\rm pQCD}(j;\xi,\eta,t,Q^2;\mu)
=
\sum_{a=\pm}
\xi^{-j}\,C_j^a\!\left(\frac{\eta}{\xi},\frac{Q^2}{\mu^2};\alpha_s\right)
H_j^a(\eta,t;\mu),
\label{eq:main-pqcd-fixedj-general}
\end{equation}
and at a conformal fixed point $\alpha_s=\alpha_s^\ast$ the Wilson coefficients become
\begin{align}
C_j^a\!\left(\frac{\eta}{\xi},\frac{Q^2}{\mu^2};\alpha_s^\ast\right)
&=
c_j^a(\alpha_s^\ast)
\left(\frac{\mu}{Q}\right)^{\gamma_j^a(\alpha_s^\ast)}
\cK_j^{(\gamma_j^a(\alpha_s^\ast))}(\vartheta),
\label{eq:main-pqcd-fixed-point-Wilson}
\end{align}
so that
\begin{align}
&\widehat{\cH}^{\rm sing}_{\rm pQCD}(j;\xi,\eta,t,Q^2;\mu)=\nonumber\\
&
\sum_{a=\pm}
\xi^{-j}\,c_j^a(\alpha_s^\ast)
\left(\frac{\mu}{Q}\right)^{\gamma_j^a(\alpha_s^\ast)}
\cK_j^{(\gamma_j^a(\alpha_s^\ast))}(\vartheta)
\left(\frac{\mu_0}{\mu}\right)^{\gamma_j^a(\alpha_s^\ast)}
H_j^a(\eta,t;\mu_0).
\label{eq:main-pqcd-fixedj}
\end{align}
The historical antecedents of this representation are the conformal-OPE constructions of Refs.~\cite{Mueller:1997hs,Belitsky:1997rh,Mueller:2005ed}, while the practically useful \DVCS synthesis is Ref.~\cite{Kumericki:2007sa}; recent NNLO extensions may be found in Refs.~\cite{Braun:2020yib,Braun:2021ysj,Braun:2022nnlo,Ji:2023transversity,Braun:2024ddvcs,Braun:2026dvcsmom}. The full derivation used here is given in Appendix~\ref{app:pqcd}.

The matching statement is now immediate. In the collinear window \eqref{eq:intro-matching-window} and at the matching point \eqref{eq:intro-matching-point}, the fixed--$j$ holographic amplitudes \eqref{eq:main-holo-fixedj} and the conformal-basis pQCD amplitudes \eqref{eq:main-pqcd-fixedj} have the same fixed--$j$ kernel-and-moment structure under the dictionary
\begin{equation}
\gamma_o(j)\longleftrightarrow \gamma_j^+(\alpha_s^\ast),
\qquad
\gamma_c(j)\longleftrightarrow \gamma_j^-(\alpha_s^\ast),
\label{eq:main-gamma-dictionary}
\end{equation}
with the nonperturbative factors identified as
\begin{subequations}
\label{eq:main-moment-dictionary}
\begin{align}
\phi_0^{o}(j,\widetilde\delta_o,\widetilde\delta_0)
\widehat d_j(\eta,t)
\widehat{\cF}_N^{(o)}(j;t)
&\longleftrightarrow
c_j^+(\alpha_s^\ast)
H_j^+(\eta,t;\mu_\ast),
\\[1ex]
\phi_0^{c}(j,\widetilde\delta_c,\widetilde\delta_0)
\widehat d_j(\eta,t)
\left(\frac{\widetilde g_5^2}{g_5^2}\right)
\widehat{\cF}_N^{(c)}(j;t)
&\longleftrightarrow
c_j^-(\alpha_s^\ast)
H_j^-(\eta,t;\mu_\ast).
\end{align}
\end{subequations}
This is the precise matching statement: a channel-by-channel identification of the fixed--$j$ hard kernel and its associated conformal moment in the stated window and at the stated scale. The arrows in Eq.~\eqref{eq:main-gamma-dictionary} identify the conformal-dimension label inside the fixed--$j$ kernel; they do not equate weak-coupling QCD anomalous-dimension functions with strong-coupling holographic trajectory functions as functions of $j$. The lower factor on the holographic side, including $\widehat d_j(\eta,t)\widehat{\cF}_N^{(X)}(j;t)$, is the holographic model representation of the conformal moment. It is not a general theorem that QCD conformal moments factorize into this particular kinematical polynomial times a $t$-form factor. The UV kernel is universal; the lower moment is matched or modeled. The remaining normalization freedom is purely conventional: changing the normalization of the UV source rescales $\phi_0^X$, while changing the normalization of the $\pm$ eigenvectors rescales $c_j^\pm$ and $H_j^\pm$ inversely. In the explicit UV convention used in Appendix~\ref{app:holo}, the cutoff factor in the boundary mode cancels against the source rescaling, so $\phi_0^X$ is a finite residual source normalization to be matched together with the forward coefficient normalization. This is the intended level of the matching.

\begin{widetext}
\begin{center}
\fbox{\begin{minipage}{0.94\textwidth}
\textbf{Fixed--$j$ claim.}  In the collinear window \eqref{eq:intro-matching-window}, after the TT scalar projection \eqref{eq:TTscalar-invariant}, and at the single matching scale \eqref{eq:intro-matching-point}, the pure-AdS upper Witten vertex produces the same hypergeometric conformal Wilson-kernel family \eqref{eq:main-universal-kernel} as the singlet conformal OPE.  The protected closed BPST branch is structurally identified with the $(-)$ conformal partial wave, the even Nishio--Watari open branch gives the unprotected $(+)$ counterpart in the projected singlet vector amplitude, and all finite source/forward normalizations and lower conformal moments are matched through \eqref{eq:main-moment-dictionary} rather than predicted by the UV kernel.
\end{minipage}}
\end{center}
\end{widetext}

The large-$N_c$ counting of the finite $c_j^\pm$ factors should also be read at the level of the products $c_j^\pm H_j^\pm$. In the unit-diagonal basis used in Appendix~\ref{app:pqcd}, $c_j^{-,(0)}$ can be $\ord(n_f/N_c)$ while $H_j^-$ carries an $\ord(N_c)$ quark admixture for even $j\ge4$; the product remains finite and normalization-convention independent. Thus an apparently small forward coefficient entry is not evidence that the closed branch is subleading; only the product with the corresponding moment has invariant meaning. This removes any apparent mismatch between a leading holographic closed branch and a suppressed forward coefficient entry.

Two immediate comments are essential. First, the matching is naturally formulated for \DDVCS, where both photons are off shell and the conformal structure is fully visible. The \DVCS limit may then be taken after the fixed--$j$ matching. Second, beyond the conformal fixed point the pure power-law scaling in $\mu$ is replaced by the familiar logarithmic LO evolution, but the same factorized logic survives. In that sense the fixed-point formula is the cleanest place to state the open/$+$ and closed/$-$ correspondence, while the running-coupling generalization tells us how the same kernel is embedded in real-QCD scale evolution.

\subsection{Scope and limitations}
\label{subsec:scope-limitations}

The result is intentionally fixed--$j$, fixed-scale, TT-projected, and kernel-level. It concerns the leading-twist unpolarized singlet vector invariant in the collinear window \eqref{eq:intro-matching-window}. At the matching scale $Q=\mu=\mu_0=\mu_\ast$, the ultraviolet Witten vertex produces the same hypergeometric conformal Wilson-kernel family as the QCD conformal OPE, while finite source/forward normalizations and lower conformal moments are matched through \eqref{eq:main-moment-dictionary}. The dictionary for $\gamma_X(j)$ identifies the conformal-dimension label inside that fixed--$j$ kernel; it is not a numerical equality between weak-coupling anomalous-dimension functions and strong-coupling trajectory functions as functions of $j$.

For phenomenology, $\mu_\ast$ should be treated as an input matching scale for the conformal moments. A fit to intermediate-$Q^2$ data should evolve the matched moments and coefficient normalizations to the measured scale, then add target-mass, higher-twist, and genuinely noncollinear corrections as corrections to the projected leading-twist description rather than as a redefinition of the fixed--$j$ $\eta/\xi$ hard kernel. The framework therefore supplies a structural matching statement and an operator-basis organization, not an all-scale equality between full theories, not a unique infrared completion, and not a global fit prescription for the full Compton tensor.

\subsection{The $j=2$ structural anchor}
\label{subsec:j2-anchor}

This is the point where the matching stops being only a kernel comparison and becomes a structural channel assignment. The first nontrivial even moment is the sharpest place to test the dictionary in Eq.~\eqref{eq:main-gamma-dictionary}. On the perturbative side, Appendix~\ref{app:pqcd} gives the LO eigen-anomalous dimensions in Eq.~\eqref{eq:app-pqcd-gamma-pm}. At $j=2$ they reduce to
\begin{equation}
\gamma_2^{-,(0)}=0,
\qquad
\gamma_2^{+,(0)}=\frac{4}{3}\left(2C_F+T_F n_f\right),
\label{eq:main-pqcd-j2-anchor}
\end{equation}
as shown explicitly in Eq.~\eqref{eq:app-pqcd-gamma-pm-at2}. The vanishing of the $(-)$ eigenvalue is the leading-order manifestation of the singlet momentum-sum rule, while the $(+)$ channel remains finite. Near $j=2$, the $(-)$ eigenvalue starts linearly from zero whereas the $(+)$ eigenvalue receives a regular finite correction, cf.\ Eq.~\eqref{eq:app-pqcd-gamma-pm-near2}. This is the perturbative side of the protected/unprotected split already emphasized in the conformal-OPE literature~\cite{Kumericki:2007sa,Braun:2026dvcsmom}.

On the holographic side, the same protected/unprotected pattern follows from the BPST assignments for the closed and open trajectories displayed in Appendix~\ref{app:holo} immediately below Eq.~\eqref{eq:delta-and-Delta}. Using Eq.~\eqref{eq:main-delta-def}, one finds for the closed channel
\begin{equation}
\Delta_c(2)=4,
\qquad
\delta_c(2)=2+\Delta_c(2)-2=4,
\qquad
\gamma_c(2)=0.
\label{eq:main-closed-j2-anchor}
\end{equation}
This is the holographic signature of the bulk graviton and hence of the conserved total energy--momentum tensor. This protected branch should not be read as a literal unmixed gluon operator at finite $N_c$; the perturbative $\pm$ eigenchannels are quark--gluon mixtures, and open/closed here denotes the effective diagonal branches of the projected large-$N_c$/holographic description. By contrast, the open branch is unprotected at the same point:
\begin{equation}
\Delta_o(2)=2+\sqrt{1+\sqrt\lambda},
\qquad
\gamma_o(2)=\Delta_o(2)-4=\sqrt{1+\sqrt\lambda}-2.
\label{eq:main-open-j2-anchor}
\end{equation}
The invariant statement is not the literal nonzero value for every possible finite $\lambda$, since the expression above has an accidental zero at $\lambda=9$. At that accidental zero the open and closed values are temporarily degenerate at $j=2$, but the open anomalous dimension remains unprotected for $j\neq2$, and the full trajectory structure in Eq.~\eqref{eq:main-branch-point-structure}, not the isolated value at one coupling, carries the unprotected assignment. The invariant statement is therefore that no Ward identity or conservation law enforces $\gamma_o(2)=0$ in the open branch. At fixed $j$, both for the physical even-spin sequence and for the complex-$j$ continuation used in Regge reconstruction, the $j=2$ match identifies the protected branch and supplies the analytic-continuation prescription used for the structural assignment:
\begin{equation}
\text{closed}\leftrightarrow (-),
\qquad
\text{open}\leftrightarrow (+).
\label{eq:main-j2-dictionary}
\end{equation}

The same conclusion is visible directly in the analytic structure of the two holographic branches. At strong coupling,
\begin{subequations}
\label{eq:main-branch-point-structure}
\begin{align}
\gamma_c(j)
&=
-j+\sqrt{4+2\sqrt\lambda\,(j-2)}
\;\simeq\;
\sqrt{2\sqrt\lambda\,(j-2)},
\\[1ex]
\gamma_o(j)
&=
-j+\sqrt{1+\sqrt\lambda\,(j-1)}
\;\simeq\;
\sqrt{\sqrt\lambda\,(j-1)}.
\end{align}
\end{subequations}
The closed trajectory is anchored by the protected spin--2 graviton point, while the open trajectory is anchored by an unprotected Reggeon branch. In the strong-coupling/intercept shorthand this appears as the familiar $\sqrt{j-2}$ versus $\sqrt{j-1}$ diagnostic. At finite $\lambda$, however, the branch points are displaced to $j_{0c}=2-2/\sqrt\lambda$ and $j_{0o}=1-1/\sqrt\lambda$; the shorthand should not be read as a statement that the finite-$\lambda$ branch points sit exactly at $2$ and $1$. The point is not that the full amplitude is saturated by $j=2$; rather, $j=2$ is the first physical even moment and therefore the cleanest diagnostic identifying the protected branch and fixing the analytic-continuation prescription used for the fixed--$j$ amplitude. Because this diagnostic rests on conservation laws and on the analytic structure of the trajectories, it is not sensitive to the particular soft-wall realization of the lower moments.

\section{Holographic factorization from the Witten diagram}
\label{sec:holo-main}

The holographic side is now derived explicitly. The strategy is simple: derive the closed channel directly from the Witten diagram, isolate the universal upper vertex, display the parallel Nishio--Watari open vertex, and use the open/closed replacement rule only as a notation-saving summary after both vertices have been written.

The closed channel is the directly derived building block because it follows from the standard $t$--channel Witten diagram and its BPST continuation~\cite{Brower:2006ea}. A key point of this paper is that the BPST closed exchange, not only the Nishio--Watari open exchange, produces the same $C_1$ hypergeometric kernel once the near-boundary $z$ powers are counted at fixed $j$. The even open channel is then written as the parallel Nishio--Watari construction of Refs.~\cite{Nishio:2014rya,Nishio:2014eua}; their Appendix~B gives the relevant open-string $C_1$ integral and hypergeometric form explicitly in Eqs.~(285)--(289). The full derivation is collected in Appendix~\ref{app:holo}; here we isolate the steps that make the matching to pQCD manifest.

\begin{figure*}[t]
\centering
\tikzset{
    photon/.style={decorate, decoration={snake, segment length=2mm, amplitude=0.5mm}, draw=black},
    boldphoton/.style={decorate, decoration={snake, segment length=2mm, amplitude=0.5mm}, draw=black, line width=0.4mm},
    particle/.style={draw=black, postaction={decorate}, decoration={markings, mark=at position .55 with {\arrow[black]{stealth}}}},
    antiparticle/.style={draw=black, postaction={decorate}, decoration={markings, mark=at position .55 with {\arrowreversed[black]{stealth}}}},
    dbl_wiggly/.style={double, decorate, decoration={snake, amplitude=1pt, segment length=3mm}, double distance=0.55pt, line width=0.8pt},
    projector/.style={densely dashed, line width=0.5pt},
    hexagram/.style={draw, star, star points=6, star point ratio=0.5, fill=black, minimum size=1.3mm, inner sep=0pt}
}

\begin{minipage}[t]{0.48\textwidth}\vspace{0pt}
\centering
\begin{tikzpicture}[scale=0.74, transform shape]
    \pgfmathsetmacro{\Rc}{sqrt(8)}

    \coordinate (P1c) at (0,0);
    \coordinate (J1c) at (0,4);
    \coordinate (P2c) at (4,0);
    \coordinate (J2c) at (4,4);

    \draw[thick] (2,2) circle[radius=\Rc cm];

    \node[hexagram, label={[align=center]below left:$\mathcal O_{P}(y_1;0)$}] (P1) at (P1c) {};
    \node[hexagram, label={[align=center]above left:$J^\mu(x_1;0)$}] (J1) at (J1c) {};
    \node[hexagram, label={[align=center]below right:$\bar{\mathcal O}_{P}(y_2;0)$}] (P2) at (P2c) {};
    \node[hexagram, label={[align=center]above right:$J^\nu(x_2;0)$}] (J2) at (J2c) {};

    \node[fill, circle, inner sep=2pt] (vb) at (2,1) {};
    \node[fill, circle, inner sep=2pt] (vt) at (2,3) {};

    \draw[particle]     (P1) -- (vb) node[pos=0.54, below, sloped] {$\Psi_{P}(p_1;z')$};
    \draw[antiparticle] (P2) -- (vb) node[pos=0.46, below, sloped] {$\bar\Psi_{P}(p_2;z')$};
    \draw[boldphoton]   (J1) -- (vt) node[pos=0.52, above, sloped] {$V(q_1;z)$};
    \draw[boldphoton]   (vt) -- (J2) node[pos=0.48, above, sloped] {$V(q_2;z)$};
    \draw[dbl_wiggly]   (vt) -- (vb) node[midway, right] {$G_j^{(c)}(q_2-q_1;z;z')$};

    \node at (2.23,2.92) {$z$};
    \node at (2.26,1.10) {$z'$};

    \node[red] at ($(J1)+(-0.48,-0.08)$) {$\frac{1}{g_5}$};
    \node[red] at ($(J2)+(0.48,-0.08)$) {$\frac{1}{g_5}$};
    \node[red] at ($(vt)+(-0.44,-0.02)$) {$\tilde{g}_5$};
    \node[red] at ($(vb)+(-0.44,0.18)$) {$\tilde{g}_5$};
\end{tikzpicture}

\vspace{0.8ex}
\refstepcounter{figure}\label{TWITTEN-2}
{\small\parbox{0.95\linewidth}{\centering
\textbf{FIG.~\thefigure.} Standard closed--channel Witten diagram before holographic collinear factorization.}}
\end{minipage}
\hfill
\begin{minipage}[t]{0.48\textwidth}\vspace{0pt}
\centering
\begin{tikzpicture}[scale=0.74, transform shape]
    \def\Rx{3.55}
    \def\Ry{3.05}

    \coordinate (J1c)  at ({\Rx*cos(136)},{\Ry*sin(136)});
    \coordinate (J2c)  at ({\Rx*cos(44)},{\Ry*sin(44)});
    \coordinate (OgLc) at ({\Rx*cos(175)},{\Ry*sin(175)});
    \coordinate (OgRc) at ({\Rx*cos(-3)},{\Ry*sin(-3)});
    \coordinate (P1c)  at ({\Rx*cos(-123)},{\Ry*sin(-123)});
    \coordinate (P2c)  at ({\Rx*cos(-57)},{\Ry*sin(-57)});

    \draw[thick] (0,0) ellipse [x radius=\Rx cm, y radius=\Ry cm];

    \node[hexagram, label={[align=center]above left:$J^\mu(x_1;0)$}] (J1) at (J1c) {};
    \node[hexagram, label={[align=center]above right:$J^\nu(x_2;0)$}] (J2) at (J2c) {};
    \node[hexagram, label={[align=center]left:$\widetilde{\mathcal O}^{(j)}_{c}(x;0)$}] (OgL) at (OgLc) {};
    \node[hexagram, label={[align=center]right:$\mathcal O^{(j)}_{c}(x';0)$}] (OgR) at (OgRc) {};
    \node[hexagram, label={[align=center]below left:$\mathcal O_{P}(y_1;0)$}] (P1) at (P1c) {};
    \node[hexagram, label={[align=center]below right:$\bar{\mathcal O}_{P}(y_2;0)$}] (P2) at (P2c) {};

    \node[fill, circle, inner sep=1.8pt] (vt) at (0.10, 1.05) {};
    \node[fill, circle, inner sep=1.8pt] (vb) at (0.15, -1.25) {};

    \draw[boldphoton] (J1) -- (vt) node[pos=0.50, above, sloped] {$V(q_1;z)$};
    \draw[boldphoton] (vt) -- (J2) node[pos=0.52, above, sloped] {$V(q_2;z)$};
    \draw[dbl_wiggly] (OgL) -- (vt) node[pos=0.48, above, sloped] {$\Psi_j^{(c),\mathrm{bdry}}(z;\varepsilon)$};

    \draw[particle]     (P1) -- (vb) node[pos=0.52, below, sloped] {$\Psi_{P}(p_1;z')$};
    \draw[antiparticle] (P2) -- (vb) node[pos=0.48, below, sloped] {$\bar\Psi_{P}(p_2;z')$};
    \draw[dbl_wiggly]   (vb) -- (OgR) node[pos=0.54, below, sloped] {$\mathcal H_j^{(c)}(K,z';\varepsilon)$};

    \draw[projector] (OgL) to[bend right=7] node[midway, below] {$\widehat d_j(\eta,t)$} (OgR);

    \node at (0.10,1.3) {$z$};
    \node at (0.16,-1.48) {$z'$};

    \node at (0,3.42) {$\mathcal C_{\gamma\gamma}^{(c)}(j)$};
    \node at (0,-3.42) {$\mathcal F_N^{(c)}(j)$};

    \node[red] at ($(J1)+(-0.38,0.04)$) {$\frac{1}{g_5}$};
    \node[red] at ($(J2)+(0.38,0.08)$) {$\frac{1}{g_5}$};
    \node[red] at ($(vt)+(-0.12,-0.25)$) {$\tilde{g}_5$};
    \node[red] at ($(vb)+(-0.25,0.18)$) {$\tilde{g}_5$};
\end{tikzpicture}

\vspace{0.8ex}
\refstepcounter{figure}\label{fig:witten-factorized-closed}
{\small\parbox{0.95\linewidth}{\centering
\textbf{FIG.~\thefigure.} Factorized closed--channel Witten diagram for holographic collinear factorization of the \DDVCS\ Compton amplitude.}}
\end{minipage}
\end{figure*}

One endpoint of the exchanged spin--$j$ propagator may be pushed to the boundary, so that the original Witten diagram factorizes into an upper three-point function and a lower hadronic matrix element. Schematically,
\begin{equation}
G_c(j,z'\to0,z;t)
\simeq
\Psi_j^{(c),\bdry}(z';\eps)
\,\cH_j^{(c)}(K,z;\eps),
\label{eq:main-schematic-factorization}
\end{equation}
where $\Psi_j^{(c),\bdry}$ is the near-boundary mode and $\cH_j^{(c)}$ the bulk-to-boundary kernel associated with the lower hadronic vertex. In pure AdS, hard-wall, and soft-wall realizations, this near-boundary factorization is the holographic counterpart of collinear factorization~\cite{Nishio:2014rya,Hatta:2018jpsi,Mamo:2019mka}. The fixed--$j$ closed-channel amplitude then takes the factorized form
\begin{equation}
\widehat{\cH}^{(c)}_{\rm holo}(j;s,t,\chi;Q_1^2,Q_2^2)
=
\bar s_c^{\,j}
\cC_{\gamma\gamma}^{(c)}(j;Q_1^2,Q_2^2,\bar\eps_c)
\widehat d_j(\eta,t)
\widehat{\cF}_N^{(c)}(j;t),
\label{eq:main-closed-factorized}
\end{equation}
with the hadronic information isolated in the bottom moment $\widehat{\cF}_N^{(c)}(j;t)$.

The universal upper vertex is obtained by taking the conformal limit of the photon bulk wave functions,
\begin{equation}
\cV^{\rm conf}(Q_i,z)=Q_i zK_1(Q_i z),
\qquad i=1,2.
\label{eq:main-conformal-photon}
\end{equation}
These are the same pure-AdS wave functions that underlie the vacuum current-current matching of Refs.~\cite{Erlich:2005qh,Grigoryan:2007vg}. The scalar kernel written below is the TT-projected upper vertex defined by Eq.~\eqref{eq:TTscalar-invariant}. In the full gauge-invariant $VVh$ or $VV\cO_j$ coupling one also generates longitudinal terms from $F_{z\mu}$ and derivative structures, which in the conformal limit may be represented through $K_0$ or $z\partial_zK_1$. After applying the scalar TT projector, these components belong to longitudinal or distinct helicity/tensor invariants rather than to the kernel matched here; the complementary LL/TL projections are the natural place where those derivative structures appear, as in Nishio--Watari and in the projection formula of Sec.~II.D, Eq.~(II.17), of Ref.~\cite{Mamo:2021tzd}. The leading scalar invariant retained here is therefore represented by $\cV(Q_1,z)\cV(Q_2,z)$. The top integral reduces to
\begin{widetext}
\begin{equation}
C_1(\delta,\eta/\xi)
\equiv
\left(1-\frac{\eta^2}{\xi^2}\right)^{1/2}
\int_0^{\infty}\dd y\,y^{1+\delta}
K_1\!\left(y\sqrt{1+\frac{\eta}{\xi}}\right)
K_1\!\left(y\sqrt{1-\frac{\eta}{\xi}}\right),
\label{eq:main-C1-def}
\end{equation}
\end{widetext}
with $y=Qz\simeq \sqrt{(Q_1^2+Q_2^2)/2}\,z$. The displayed square-root form is the spacelike \DDVCS domain $|\eta/\xi|<1$ in which both virtualities are kept off shell; endpoint limits such as \DVCS and complex physical continuations are taken after the fixed--$j$ kernel has been identified. The integral is elementary:\footnote{Equivalently, Eq.~\eqref{eq:main-C1-result} is the standard product-Bessel integral for $K_1(a y)K_1(b y)$ after setting $a^2=1+\eta/\xi$ and $b^2=1-\eta/\xi$. Nishio--Watari use the same identity for the open-string $C_1$ integral in Appendix~B of Ref.~\cite{Nishio:2014eua}.}
\begin{equation}
C_1(\delta,\eta/\xi)
=
2^{\delta-1}\frac{\delta+2}{\delta}
\frac{\Gamma\!\left(\frac{\delta}{2}+1\right)^4}{\Gamma(\delta+2)}
{}_2F_1\!\left(
\frac{\delta}{4},\frac{\delta}{4}+\frac12;
\frac{\delta}{2}+\frac32;
\frac{\eta^2}{\xi^2}
\right).
\label{eq:main-C1-result}
\end{equation}
The hard kernel is therefore universal. It is not a model artifact of the soft-wall; it is fixed completely by the pure-$\AdS$ photon wave functions and by the power of $z$ carried by the exchanged spin--$j$ field. For the closed BPST branch this is the new part of the construction: the same hypergeometric family arises from a closed-string Witten diagram with its Mellin exponent fixed by BPST power counting. This is the precise mathematical sense in which the hard kernel is a consequence of the AdS geometry rather than of a phenomenological ansatz.

The exponent entering Eq.~\eqref{eq:main-C1-def} is not assumed. It follows directly from the Witten diagram. In the upper vertex the measure contributes $z^{-5}$, the tensor vertex contributes $z^{4+2(j-2)}$, the bulk-to-boundary propagator contributes $z^{-(j-2)}$, the boundary mode contributes $z^{\Delta_X(j)}$, and the two photon wave functions contribute $z^2$. Hence the integrand scales as
\begin{equation}
z^{-5}\times z^{4+2(j-2)}\times z^{-(j-2)}\times z^{\Delta_X(j)}\times z^2
=
z^{j+\Delta_X(j)-1}
=
z^{1+\delta_X(j)},
\label{eq:main-z-counting}
\end{equation}
which fixes the Mellin exponent to Eq.~\eqref{eq:main-delta-def}. This $z$-power counting is the structural anchor on the holographic side: it is what turns the Witten diagram into the same anomalous-dimension combination that appears in the conformal-basis Wilson coefficients of pQCD.

All infrared model dependence is confined to the lower vertex. In the soft-wall model the right- and left-handed nucleon contributions reduce to gamma-function expressions, and the bottom conformal moment is
\begin{equation}
\widehat{\cF}_N^{(X)}(j;t)=\frac12\Big[\widehat{\cF}_R^{(X)}(j;t)+\widehat{\cF}_L^{(X)}(j;t)\Big],
\qquad X=o,c,
\label{eq:main-bottom-moment-average}
\end{equation}
with the explicit formulas collected in Appendix~\ref{app:holo}. The important point is not the particular soft-wall expression, but the separation itself: the hard upper factor is universal, while the hadronic vertex carries the entire infrared completion. In this sense the soft-wall model is used here as an input for the lower conformal moments, not as the source of the Wilson-kernel-like factor.

Before using the compact map, it is helpful to display the corresponding open upper Witten vertex explicitly. In the even Nishio--Watari open channel one has
\begin{widetext}
\begin{equation}
\widetilde{\cC}_{\gamma\gamma}^{(o)}(j;Q_1^2,Q_2^2,\beps_o)
=
\frac{g_5}{2g_5^2}
\int_0^{\infty}\dd z\,\sqrt g\,\ee^{-\Phi_o}
\,z^{4+2(j-2)}
\cK_{\gamma\gamma}^{(o)}(Q_1,Q_2;z)
\,z^{-(j-2)}\Psi_j^{(o),\bdry}(z;\eps),
\label{eq:main-open-top-vertex-preconf}
\end{equation}
with
\begin{equation}
\cK_{\gamma\gamma}^{(o)}(Q_1,Q_2;z)=\cV(Q_1,z)\cV(Q_2,z),
\qquad
\Psi_j^{(o),\bdry}(z;\eps)
=
-\frac{(\sqrt2\kappa_o z)^{\Delta_o(j)}(\sqrt2\beps_o)^{4-\Delta_o(j)}}{\Delta_o(j)}.
\label{eq:main-open-boundary-mode}
\end{equation}
\end{widetext}
Taking \(\cV(Q_i,z)\to Q_i zK_1(Q_i z)\), the same \(z\)-power counting gives \(z^{j+\Delta_o(j)-1}=z^{1+\delta_o(j)}\), and hence the same \(C_1(\delta_o,\eta/\xi)\) integral and the same kernel \(\cK_j^{(\gamma_o)}(\vartheta)\). Thus the open result is not asserted from the closed result; it is the parallel Nishio--Watari open-string Witten vertex written in the same fixed--\(j\) factorized form. Appendix~B of Ref.~\cite{Nishio:2014eua} displays this $C_1$ integral and its hypergeometric form in Eqs.~(285)--(289).

Once the open vertex has been displayed, the even open channel can be summarized by the replacement rule
\begin{widetext}
\begin{equation}
\widetilde g_5\rightarrow g_5,
\qquad
\kappa_c\rightarrow \kappa_o,
\qquad
h_{\mu\nu}^{c}\rightarrow h_{\mu\nu}^{o},
\qquad
\Delta_c\rightarrow \Delta_o,
\qquad
\gamma_c\rightarrow \gamma_o.
\label{eq:main-replacement-rule}
\end{equation}
\end{widetext}
This rule is a shorthand after the two projected vertices have been independently specified; it is not a claim about the full open-string tensor sector. It is used here at the level of the projected upper scalar kernel and the resulting fixed--$j$ invariant amplitude. In the Nishio--Watari construction, the same current insertions and the same spin--$j$ kinematical projector couple to an open-string Regge trajectory with different open-string boundary data, couplings, and intercept~\cite{Nishio:2014rya,Nishio:2014eua}. Moreover, graviton-like effective descriptions of \(2^{++}\) tensor mesons have long been used in AdS/QCD~\cite{Katz:2005ir}, so employing a parallel spin-2 action on the open side is not foreign to holographic hadron phenomenology. The continued phenomenological relevance of tensor-meson exchange in modern hadronic amplitudes provides an additional supporting perspective~\cite{Mager:2025pvz}. Once the upper vertex is fixed by the pure-AdS photon wave functions, the only channel dependence compatible with that construction enters through the substitutions in Eq.~\eqref{eq:main-replacement-rule}. The open and closed channels therefore have the same functional dependence in the universal kernel, differing only in the trajectory data, the overall coupling factor, and the lower hadronic moments. This is precisely the structure needed for the open/$+$ and closed/$-$ correspondence.

\section{pQCD conformal OPE and fixed--$j$ structural matching}
\label{sec:pqcd-main}

With the holographic side in hand, the remaining task is to write the perturbative amplitude in the same fixed--$j$ language and match the two structures term by term.

The perturbative side of the comparison is most transparent in the conformal basis that diagonalizes the LO singlet evolution. Historically, this representation grows out of the conformal-OPE program of Refs.~\cite{Mueller:1997hs,Belitsky:1997rh}, the complex conformal-spin construction of Ref.~\cite{Mueller:2005ed}, and the explicit LO evolution solutions of Refs.~\cite{Kirch:2005hu,Manashov:2005qm}. For \DVCS phenomenology and conventions, however, the practically most useful anchor is Ref.~\cite{Kumericki:2007sa}, especially Sec.~4.1, which is the basis followed here. Denote the corresponding conformal moments by $H_j^\pm(\eta,t;\mu)$. At a conformal fixed point the fixed--$j$ Wilson coefficients are given by Eq.~\eqref{eq:main-pqcd-fixed-point-Wilson}, and the full singlet fixed--$j$ amplitude is Eq.~\eqref{eq:main-pqcd-fixedj}. The key fact is that the same Gauss hypergeometric function appears on both sides, with the same combination of spin and anomalous dimension entering its parameters.

This agreement is not automatic merely because both descriptions know about the conformal group. Conformal symmetry fixes the possible three-point functional form once the spin and dimension of the exchanged operator are specified, but the nontrivial content here is the following three-part statement:
\begin{enumerate}
\item the Witten diagram factorizes at fixed $j$ into an upper current kernel and a lower conformal moment;
\item the upper vertex fixes the specific Mellin label $\delta_X(j)=j+\Delta_X(j)-2=2j+\gamma_X(j)$ through the $z$-power count in Eq.~\eqref{eq:main-z-counting}, rather than by assumption;
\item the protected/unprotected $j=2$ diagnostic identifies the closed protected branch and supplies the analytic-continuation prescription for the closed/open $\leftrightarrow$ $(-)/(+)$ channel assignment used in this projected amplitude.
\end{enumerate}
The shared conformal group explains why a hypergeometric kernel can appear; these three Witten-diagram and channel facts explain why the particular kernel and dictionary are the ones matched here.

The matching can therefore be stated without ambiguity. In the collinear window and at the matching point $Q=\mu=\mu_0=\mu_\ast$,
\begin{subequations}
\label{eq:main-channel-matching}
\begin{align}
\widehat{\cH}^{(o)}_{\rm holo}(j)\Big|_{\mu=Q=\mu_0=\mu_\ast}
&\longleftrightarrow
\xi^{-j}
c_j^+(\alpha_s^\ast)
\cK_j^{(\gamma_j^+(\alpha_s^\ast))}(\vartheta)
H_j^+(\eta,t;\mu_\ast),
\\[1.5ex]
\widehat{\cH}^{(c)}_{\rm holo}(j)\Big|_{\mu=Q=\mu_0=\mu_\ast}
&\longleftrightarrow
\xi^{-j}
c_j^-(\alpha_s^\ast)
\cK_j^{(\gamma_j^-(\alpha_s^\ast))}(\vartheta)
H_j^-(\eta,t;\mu_\ast).
\end{align}
\end{subequations}
In this sense the open-string contribution provides the fixed--$j$ holographic realization of the $(+)$ channel, while the closed-string contribution provides the corresponding realization of the $(-)$ channel in the projected singlet vector amplitude. These equations are the core result. They state that at fixed $j$, in the window \eqref{eq:intro-matching-window} and at the single scale \eqref{eq:intro-matching-point}, the two descriptions have the same kernel-and-moment structure under the stated dictionary.

The $j=2$ anchor makes this statement sharper. Equation~\eqref{eq:main-pqcd-j2-anchor} shows that the perturbative $(-)$ eigenvalue is the protected one, while Eqs.~\eqref{eq:main-closed-j2-anchor} and \eqref{eq:main-open-j2-anchor} show that the protected holographic branch is the closed channel and that the open branch has no symmetry-enforced zero at the same point. The analytic behavior in Eq.~\eqref{eq:main-branch-point-structure} then extends this protected/unprotected distinction away from $j=2$. The channel assignment is therefore anchored before any resummation or phenomenological fit is considered.

The detailed pQCD derivation matters for three reasons. First, the constants $c_j^\pm$ are not trivial in the normalization convention in which the diagonalization matrix has unit diagonal entries. They are fixed by matching to the forward DIS Wilson coefficients, not by guesswork; Appendix~\ref{app:pqcd} makes this explicit in Eqs.~\eqref{eq:app-pqcd-cpm-general}--\eqref{eq:app-pqcd-cpm-LO-explicit}. Second, the normalization freedom of the $\pm$ eigenvectors mirrors the source-normalization freedom on the holographic side. Third, once those conventions are fixed, the correspondence between Eqs.~\eqref{eq:main-holo-fixedj} and \eqref{eq:main-pqcd-fixedj} is a statement about the physical fixed--$j$ amplitude itself rather than a statement with free normalization on the perturbative side.

It is equally important to state what the conformal fixed-point formulas do and do not assume. They provide the cleanest analytic representation of the hard kernel, but the matching itself is local in scale. At any chosen scale $\mu_\ast$, one may evaluate both descriptions at
\begin{equation}
Q=\mu=\mu_0=\mu_\ast,
\end{equation}
so that the fixed--$j$ pQCD amplitude takes the schematic form
\begin{widetext}
\begin{equation}
\widehat{\cH}^{\rm sing}_{\rm pQCD}(j)\Big|_{\mu=Q=\mu_0=\mu_\ast}
=
\sum_{a=\pm}
\xi^{-j}\,c_j^a(\alpha_s(\mu_\ast))
\,\cK_j^{(\gamma_j^a(\alpha_s(\mu_\ast)))}(\vartheta)
\,H_j^a(\eta,t;\mu_\ast),
\label{eq:main-single-scale-match}
\end{equation}
\end{widetext}
where the displayed $\cK_j^{(\gamma_j^a)}$ is the same hypergeometric kernel with the anomalous dimension evaluated at the chosen scale. In this form the matching is an ordinary effective-theory matching at one scale. In the conformal partial-wave/CS representation of Ref.~\cite{Kumericki:2007sa}, the object matched here is the fixed--$j$ conformal kernel of the projected invariant. Conformal-breaking, beta-proportional terms, and finite scheme transformations are assigned to coefficient/evolution bookkeeping away from $Q=\mu=\mu_0=\mu_\ast$; they do not define a second fixed-scale $\eta/\xi$ kernel for this projection. Higher-order perturbative information enters through $\gamma_j^a(\alpha_s(\mu_\ast))$ and $c_j^a(\alpha_s(\mu_\ast))$; the remaining conformal-anomaly and running-coupling bookkeeping governs the evolution away from the matching point.

Away from a conformal fixed point the coupling runs. The pure power laws in $\mu$ are then replaced by the usual logarithmic evolution factors,
\begin{equation}
\mathcal E_j^\pm(\mu,\mu_0)
=
\left[\frac{\alpha_s(\mu)}{\alpha_s(\mu_0)}\right]^{\gamma_j^{\pm,(0)}/\beta_0},
\label{eq:main-running-evolution}
\end{equation}
and the Wilson coefficients admit the usual LO RG improvement. The explicit formulas are given in Appendix~\ref{app:pqcd}, especially Eqs.~\eqref{eq:app-pqcd-LO-evolution} and \eqref{eq:app-pqcd-final-LO-running}. The main point is that the same factorized logic survives: logarithmic running changes the scale dependence and the bookkeeping of moments and coefficients, not the fixed-scale hypergeometric functional form of the hard kernel or the open/$+$, closed/$-$ assignment. This is why the fixed-point formulas should be read as a presentational convenience, not as a hidden dynamical assumption.

Recent NNLO results for \DVCS and \DDVCS coefficient functions sharpen the perturbative side of the story~\cite{Braun:2020yib,Braun:2021ysj,Braun:2022nnlo,Ji:2023transversity,Braun:2024ddvcs,Braun:2026dvcsmom}. They do not change the scope of the present structural matching claim. Rather, they refine the perturbative data entering the same conformal partial wave: the anomalous dimensions $\gamma_j^\pm(\alpha_s)$, the finite coefficient normalizations $c_j^\pm(\alpha_s)$, and the evolution bookkeeping. The fixed-scale hypergeometric functional dependence in $\eta/\xi$ is the kernel being matched.

\section{Physical amplitude, operator meaning, and phenomenological scope}
\label{sec:physical-meaning}

Having established the fixed--$j$ dictionary, we now return to the physical amplitude and to the question of what this structure means beyond the formal matching itself.

The fixed--$j$ amplitudes are the building blocks of the physical \DDVCS Compton form factor. Polynomiality is a statement about the physical even-spin local moments; the complex-$j$ representation below is the analytic continuation of that sequence used for Regge reconstruction. The reggeized amplitude is obtained by the even-signature Sommerfeld--Watson transform,
\begin{widetext}
\begin{equation}
\widehat{\cH}(\bar s,\bar t,\chi;Q_1^2,Q_2^2)
=
\sum_{X=o,c}
\int_{\cC_X}\frac{\dd j}{2\pi\ii}
\,\xi_+(j)\,\widehat{\cH}^{(X)}_{\rm holo}(j),
\qquad
\xi_+(j)=\frac{1+\ee^{-\ii\pi j}}{\sin \pi j},
\label{eq:main-SW}
\end{equation}
\end{widetext}
with only even spins contributing in the sector considered here.

After holographic collinear factorization the operator interpretation is immediate. The upper vertex is the universal three-point function of two electromagnetic currents and a normalized spin--$j$ singlet operator in the corresponding channel, while the lower vertex is the matrix element of that same operator between nucleon states,
\begin{widetext}
\begin{equation}
\text{upper vertex}
\sim \langle J^\mu J^\nu\widetilde{\cO}^{(j)}_X\rangle,
\qquad
\text{lower vertex}
\sim \langle P'|\cO_X^{(j)}|P\rangle,
\qquad X=o,c.
\label{eq:main-operator-interpretation}
\end{equation}
\end{widetext}
The universal hypergeometric function is therefore the holographic analogue of the perturbative Wilson coefficient, whereas the bottom moments $\widehat{\cF}_N^{(X)}$ are the holographic analogue of the nonperturbative conformal moments.

The same statement also clarifies the role of the conformal/Gegenbauer basis in phenomenology. An arbitrary complete orthogonal basis can represent a function, but it need not diagonalize evolution, correspond to local twist-two operator matrix elements, or make polynomiality of the skewness expansion transparent. The conformal basis does all three. In the present holographic construction it has a fourth property: its moments are precisely the quantities represented by the lower fixed--\(j\) Witten vertex. This is why working with conformal moments is a physically meaningful way to attack the deconvolution problem in \DVCS/\DDVCS extraction, rather than a cosmetic re-expansion of the same information.

This conceptual organization gives a controlled operator interpretation for phenomenology. Once the universal kernel is fixed, the nonperturbative information resides in conformal moments that may be modeled, extracted, or compared across frameworks, including lattice-QCD determinations of local moments and form factors. This is also the sense in which the present analysis bridges the high-energy Regge picture and the collinear-factorization picture: the former provides the reggeized continuation, while the latter fixes the operator language in which the fixed--$j$ amplitude is interpreted.

For phenomenological use, the practical prescription is therefore simple. Keep the fixed--$j$ kernel $\cK_j^{(\gamma)}(\eta/\xi)$ fixed, choose or fit the lower moment data $\widehat{\cF}_N^{(o,c)}(j;t)$ equivalently mapped to $H_j^\pm(\eta,t;\mu_\ast)$, impose polynomiality on the physical even-spin moments, and then reconstruct the Compton form factor by the Mellin-Barnes/Sommerfeld--Watson representation with the usual QCD evolution away from $\mu_\ast$. A concrete numerical implementation would choose a soft-wall or alternative IR ansatz for $\widehat{\cF}_N^{(o,c)}$, fix the finite factors $c_j^\pm(\alpha_s(\mu_\ast))$ at the matching scale, evolve to the experimental scale, and then predict or fit channel ratios such as $c_j^+H_j^+/(c_j^-H_j^-)$ at given $(\eta,t)$. At intermediate experimental $Q^2$, target-mass, higher-twist, and noncollinear corrections should be included as additional terms or systematic uncertainties. In such a fit the adjustable quantities are the conformal-moment parameters and infrared scales, not the $\eta/\xi$ hard kernel.

The role of the $j=2$ anchor should be read in this way. The first even moment is an anchor, not a saturation mechanism for the full reggeized amplitude: it fixes the protected versus unprotected channel assignment and therefore selects the operator dictionary that organizes the full fixed--$j$ amplitude. Once that dictionary is established, the Sommerfeld--Watson reconstruction in Eq.~\eqref{eq:main-SW} is no longer a model-dependent guess about which channel is which. It is the continuation of a channel assignment already anchored by symmetry and by the analytic structure of the two branches.

The \DVCS limit may be taken only after the fixed--$j$ matching has been established in \DDVCS. Technically, one continues one external virtuality to zero after factorization and after identifying the fixed--$j$ kernel. No new hard kernel is introduced in that limit; what changes is that the fully off-shell conformal visibility of the \DDVCS kinematics is reduced. In the Bjorken-like \DVCS limit one has \(\eta/\xi\to 1\), so the Gauss hypergeometric function appearing in the universal kernel reduces by Gauss' theorem to a pure gamma-function ratio,
\begin{equation}
{}_2F_1\!\left(
\frac{\delta}{4},\frac{\delta}{4}+\frac12;\frac{\delta}{2}+\frac32;1
\right)
=
\frac{\Gamma\!\left(\frac{\delta}{2}+\frac32\right)}
{\Gamma\!\left(\frac{\delta}{4}+1\right)\Gamma\!\left(\frac{\delta}{4}+\frac32\right)},
\label{eq:main-dvcs-limit-hypergeometric}
\end{equation}
since \(c-a-b=1\) for the parameters in Eq.~\eqref{eq:main-C1-result}. Thus the same universal kernel survives, but its dependence on the off-forward ratio \(\eta^2/\xi^2\) collapses to its on-shell endpoint value. Together with the forward-skewness check \(\cK_j^{(\gamma)}(0)=1\), this gamma-function endpoint is the \DVCS counterpart of the same conformal partial wave whose forward normalization fixes $c_j^\pm$; it is not a new Wilson-kernel family. This is why \DDVCS is the natural arena for the structural matching statement, while \DVCS is the phenomenologically familiar descendant.

\subsection{Useful fixed--\texorpdfstring{$j$}{j} checks}
\label{subsec:fixedj-checks}

Several simple limits make the structure transparent. First, in the forward-skewness limit,
\begin{equation}
\cK_j^{(\gamma)}(0)=1,
\label{eq:check-forward}
\end{equation}
so the nonforward Wilson kernel reduces to its forward normalization \(c_j^\pm\), as required by the DIS normalization used in Appendix~\ref{app:pqcd}. Second, in the tree-level or zero-anomalous-dimension limit,
\begin{equation}
\cK_j^{(0)}(\vartheta)
=
{}_2F_1\!\left(\frac j2,\frac{j+1}{2};j+\frac32;\vartheta^2\right),
\label{eq:check-gamma-zero}
\end{equation}
which is the standard conformal partial wave of the leading vector OPE. Third, the first physical even spin tests the channel dictionary:
\begin{equation}
\gamma_c(2)=0\leftrightarrow \gamma_2^{-,(0)}=0,
\qquad
\gamma_o(2)\ \text{unprotected}\leftrightarrow \gamma_2^{+,(0)}\neq0,
\label{eq:check-j2}
\end{equation}
up to the fixed-scale dictionary for the anomalous-dimension label. Here ``unprotected'' means that no conservation law enforces a zero in the open branch; in the large-$\lambda$ regime the value is finite and positive. Finally, the \DVCS endpoint is precisely Eq.~\eqref{eq:main-dvcs-limit-hypergeometric}. These four checks show that no extra skewness kernel is being hidden in the matching; the same hypergeometric family controls the forward, conformal, protected-spin, and one-real-photon limits.

The same perspective also clarifies the role of model dependence. The pure-$\AdS$ upper vertex does not know about the soft-wall, nor about any other infrared completion. Changing the infrared model changes the lower moments $\widehat{\cF}_N^{(X)}(j;t)$, but it does not change the universal hypergeometric kernel. In that sense the soft-wall model used here is a definite choice for the hadronic conformal moments, not a choice that affects the hard factor. The fixed--$j$ structural matching to the conformal-basis Wilson-kernel family is therefore a statement about the structure of the holographic framework itself, not about a particular infrared ansatz.

Finally, the open/$+$ and closed/$-$ dictionary explains why holographic QCD naturally packages several older hadronic ideas into a single constrained framework. Regge behavior, vector-meson dominance, large-$N_c$ organization, and Pomeron-like exchange are correlated by the same fixed--$j$ factorization structure that is now seen to match the conformal-basis organization of the QCD amplitude. In that sense the present result is a concrete bridge between the holographic Witten-diagram description and the fixed--$j$ factorized structure of the singlet conformal OPE. It is also the natural hadronic-state extension of the more elementary holographic/QCD matching seen in vacuum current-current correlators~\cite{Erlich:2005qh,Grigoryan:2007vg}, see also the companion Letter \cite{Mamo:2026vuq}.

\section{Conclusion}
\label{sec:conclusion}

We have reformulated the holographic description of \DDVCS as a fixed--$j$ structural matching problem. The result of the paper has three parts. First, the holographic fixed--$j$ amplitude factorizes into a universal upper kernel and a lower hadronic conformal moment. Second, the universal upper kernel is the same Gauss hypergeometric function that appears in the conformal-basis Wilson coefficients of the singlet vector Compton form factor in QCD. Third, the first physical even moment $j=2$ identifies the protected closed branch and supplies the analytic-continuation prescription for the closed/open \(\leftrightarrow\) $(-)/(+)$ channel assignment: the closed trajectory is anchored by the protected spin--2 graviton point, while the open trajectory is anchored by an unprotected Reggeon branch. In the strong-coupling/intercept shorthand this appears as the familiar $\sqrt{j-2}$ versus $\sqrt{j-1}$ diagnostic, with the finite-$\lambda$ branch points displaced as stated above.

The distinction from Nishio--Watari is central. Their open-string analysis already contains the $C_1$ integral and its hypergeometric form. The closed channel here is the new rigorous starting point: its BPST near-boundary factorization isolates a universal upper photon vertex and derives the same hypergeometric family with the precise Mellin exponent $\delta_c(j)=j+\Delta_c(j)-2$ fixed by $z$-power counting. Put tersely, the novelty is not the open \(C_1\) identity itself; it is that the closed BPST fixed--\(j\) upper Witten vertex independently produces the same conformal Wilson-kernel family, fixes the Mellin label by \(z\)-power counting, and thereby completes the structural open/closed \(\leftrightarrow(+)/(-)\) operator-basis mapping relevant for \DDVCS/\DVCS deconvolution. The even open channel is displayed through its corresponding Nishio--Watari open Witten vertex and then summarized by a parallel replacement rule. The new use of the open channel here is not the rediscovery of its $C_1$ integral; it is its placement in the modern $\pm$ eigenbasis with explicit $c_j^\pm$ normalization bookkeeping and with the $j=2$ diagnostic supplying the $(+)$ branch prescription. In both channels the upper impact factor is fixed by the pure-$\AdS$ photon wave functions and gives a closed-form Gauss hypergeometric function of $\eta^2/\xi^2$, with Mellin exponent $\delta_X(j)=j+\Delta_X(j)-2=2j+\gamma_X(j)$. This is the point at which the holographic construction stops looking like ordinary AdS/QCD phenomenology and becomes a controlled bridge to the conformal-OPE language.

The role of the conformal fixed-point formulas should be read precisely. They provide the cleanest analytic representation of the hard kernel, but the matching itself is imposed at the single scale \eqref{eq:intro-matching-point} in the collinear window \eqref{eq:intro-matching-window}. Away from the fixed point, LO running replaces the pure power laws by the familiar logarithmic evolution factors but does not alter the fixed-scale projected kernel-level statement or the channel assignment. The matching statement is therefore not a claim of all-scale equality between full theories. It is a fixed--$j$ channel-by-channel identification for the hard kernel and its associated conformal moment. Recent NNLO \DVCS/\DDVCS coefficient functions refine the perturbative input entering the same conformal partial wave---$\gamma_j^\pm(\alpha_s)$, $c_j^\pm(\alpha_s)$, and evolution terms---rather than replacing the fixed-scale hypergeometric kernel~\cite{Braun:2020yib,Braun:2021ysj,Braun:2022nnlo,Ji:2023transversity,Braun:2024ddvcs,Braun:2026dvcsmom}.

The main implication is straightforward. Once organized at fixed $j$, holographic \DDVCS/\DVCS realizes the same conformal-basis hard-kernel structure that underlies collinear factorization, while encoding the hadronic information in nonperturbative conformal moments that can be compared to other nonperturbative determinations, including lattice-QCD moments and form factors. This gives a concrete physics reason why the conformal/Gegenbauer basis is special compared with an arbitrary orthogonal expansion: the same basis simultaneously diagonalizes the leading singlet operator problem, enforces skewness polynomiality through local conformal moments, and admits a geometric holographic realization of those moments through lower Witten vertices. In that precise sense the present framework provides a controlled and physically interpretable language for hadronic Compton scattering, and a basis-level way to organize the \DVCS/\DDVCS deconvolution problem emphasized in Ref.~\cite{Bertone:2021yyz}, rather than a disconnected phenomenological ansatz. Outside the stated collinear matching window, higher-twist effects, target-mass corrections, and genuinely noncollinear infrared dynamics can modify this simple fixed--$j$ universality; those lie beyond the present structural claim.

\section*{Acknowledgments}
I thank Christian Weiss for discussions.  I also thank Jefferson Lab for hospitality during the completion of this work.  This work was supported by DOE grant no. DE-FG02-04ER41309, NSF grant no. 2412625, and DOE under the umbrella of the Quark-Gluon Tomography (QGT) Topical Collaboration with Award No. DE-SC0023646.

\appendix

\begin{widetext}
\section{Detailed holographic derivation}
\label{app:holo}
\subsection{Kinematics, soft--wall setup, and canonical normalization}
\label{sec:setup}

We work in the Poincar\'e patch of $\AdS_5$,
\begin{equation}
\dd s^2
=
\frac{R^2}{z^2}\left(\eta_{\mu\nu}\dd x^\mu\dd x^\nu-\dd z^2\right),
\qquad
\eta_{\mu\nu}=\diag(1,-1,-1,-1),
\end{equation}
with sector--dependent soft--wall dilaton profiles
\begin{equation}
\Phi_o(z)=\kappa_o^2 z^2,
\qquad
\Phi_c(z)=\kappa_c^2 z^2,
\end{equation}
and we set $R=1$ from now on.  The external transverse photon kernel satisfies
\begin{equation}
\partial_z\left(\frac{\ee^{-\Phi_o}}{z}\partial_z\cV(Q_1,z)\right)
-
\frac{Q_1^2\ee^{-\Phi_o}}{z}\cV(Q_1,z)=0,
\qquad
\cV(Q_1,\eps)=1,
\end{equation}
with solution
\begin{equation}
\cV(Q_1,z)
=
\kappa_o^2 z^2
\Gamma\!\left(1+\frac{Q_1^2}{4\kappa_o^2}\right)
U\!\left(1+\frac{Q_1^2}{4\kappa_o^2},2;\kappa_o^2 z^2\right).
\end{equation}
For the upper vertex we will later take the conformal limit,
\begin{equation}
\cV(Q_1,z)\longrightarrow \cV^{\rm conf}(Q_1,z)=Q_1zK_1(Q_1z),
\end{equation}
because the hard upper impact factor is insensitive to the IR completion.  That is precisely why
the upper vertex is universal.

For the twist--$\tau$ nucleon ground state we use the soft--wall wave functions
\begin{subequations}
\begin{align}
\psi_R(z;0)
&=
\frac{\wt n_R}{\kappa_o^{\tau-2}}z^{5/2}(\kappa_o^2z^2)^{(\tau-2)/2},
\\
\psi_L(z;0)
&=
\frac{\wt n_L}{\kappa_o^{\tau-1}}z^{5/2}(\kappa_o^2z^2)^{(\tau-1)/2},
\\
\wt n_L
&=
\kappa_o^{\tau}\sqrt{\frac{2}{\Gamma(\tau)}},
\qquad
\wt n_R
=
\wt n_L\,\kappa_o^{-1}\sqrt{\tau-1}.
\end{align}
\end{subequations}

The symmetric closed sector is described by the graviton--like fluctuation $h_{\mu\nu}^{c}$ with
quadratic action
\begin{equation}
S_c^{(2)}
=
\frac{1}{R}
\int \dd^4x\,\dd z\,\sqrt g\,\ee^{-2\Phi_c}
\left[
-\frac{1}{4\wt g_5^2}
\,g^{MN}\eta^{\lambda\rho}\eta^{\sigma\tau}
\partial_M h^{c}_{\lambda\sigma}\partial_N h^{c}_{\rho\tau}
+\frac{1}{8\wt g_5^2}
\,g^{MN}\eta^{\alpha\beta}\eta^{\gamma\sigma}
\partial_M h^{c}_{\alpha\beta}\partial_N h^{c}_{\gamma\sigma}
\right],
\end{equation}
where $\wt g_5^2=2\kappa_5^2$ and axial gauge $h_{\mu z}^{c}=h_{zz}^{c}=0$ is understood.  The
even--spin open sector is described by the parallel graviton--like field $h_{\mu\nu}^{o}$,
\begin{equation}
S_o^{(2)}
=
\frac{1}{R}
\int \dd^4x\,\dd z\,\sqrt g\,\ee^{-2\Phi_o}
\left[
-\frac{1}{4g_5^2}
\,g^{MN}\eta^{\lambda\rho}\eta^{\sigma\tau}
\partial_M h^{o}_{\lambda\sigma}\partial_N h^{o}_{\rho\tau}
+\frac{1}{8g_5^2}
\,g^{MN}\eta^{\alpha\beta}\eta^{\gamma\sigma}
\partial_M h^{o}_{\alpha\beta}\partial_N h^{o}_{\gamma\sigma}
\right],
\end{equation}
which is the closed action with $\wt g_5\to g_5$, $\Phi_c\to\Phi_o$, and
$h_{\mu\nu}^{c}\to h_{\mu\nu}^{o}$.  The field $h_{\mu\nu}^{o}$ is used here as an effective projected even-spin open Reggeon/tensor exchange in the scalar invariant, not as a complete open-string tensor action.

The canonically normalized bulk fluctuations are therefore
\begin{equation}
\Psi\rightarrow g_5\Psi,
\qquad
V_M\rightarrow g_5V_M,
\qquad
h_{\mu\nu}^{o}\rightarrow g_5 h_{\mu\nu}^{o},
\qquad
h_{\mu\nu}^{c}\rightarrow \wt g_5 h_{\mu\nu}^{c}.
\end{equation}
Each external non--normalizable photon source contributes $1/g_5$, each open cubic vertex
contributes $g_5$, and each closed cubic vertex contributes $\wt g_5$.  Those are the coupling
factors shown in red in Figs.~\ref{TWITTEN-2} and \ref{fig:witten-factorized-closed}.

The Regge trajectories are written as
\begin{equation}
\Delta_X(j)=2+j+\gamma_X(j),
\qquad
\delta_X(j)=j+\Delta_X(j)-2=2j+\gamma_X(j),
\qquad X=o,c.
\label{eq:delta-and-Delta}
\end{equation}
For the closed branch we use the BPST form
\begin{equation}
\Delta_c(j)=2+\sqrt{2\sqrt\lambda\,(j-j_{0c})},
\qquad
j_{0c}=2-\frac{2}{\sqrt\lambda},
\end{equation}
while the even open branch is taken to be
\begin{equation}
\Delta_o(j)=2+\sqrt{\sqrt\lambda\,(j-j_{0o})},
\qquad
j_{0o}=1-\frac{1}{\sqrt\lambda}.
\end{equation}
The rest of the derivation is organized so that the closed channel is derived directly and the open channel is
introduced through the parallel Nishio--Watari vertex before the replacement rule \eqref{eq:main-replacement-rule} is used as shorthand. Also note that at strong 't Hooft coupling $\lambda\gg1$ and fixed-$j$, the closed BPST branch can be rewritten as
\begin{align}
\gamma_{c}(j)=\Delta_c(j)-2-j=-j+\sqrt{4+2\sqrt\lambda\,(j-2)}\simeq \sqrt{2\sqrt\lambda\,(j-2)}\,,
\end{align}
which vanishes at $j=2$. Similarly, the open BPST branch can be rewritten as
\begin{align}
\gamma_{o}(j)=\Delta_o(j)-2-j=-j+\sqrt{1+\sqrt\lambda\,(j-1)}\simeq \sqrt{\sqrt\lambda\,(j-1)}\,,
\end{align}
which is finite and positive at $j=2$ in the large-$\lambda$ regime. At finite $\lambda$, the important structural point is that this open-branch value is not protected by a conservation law; an accidental zero at a special finite coupling would only create a local degeneracy at $j=2$ and would not change the open/closed branch prescription away from that point.

\subsection{Closed channel: from the standard Witten diagram to holographic collinear factorization}
\label{sec:closed-factorization}

The closed channel is the directly derived building block because it follows from the standard
soft--wall graviton Witten diagram.  The first step is the spectral decomposition of the spin--$j$
propagator.

\subsection{Spin--\texorpdfstring{$j$}{j} modes, propagator, and near--boundary factorization}

The normalizable closed modes are
\begin{equation}
\psi_n^{(c)}(j,z)
=
 c_n^{(c)}(j)
 z^{\Delta_c(j)}
 L_n^{\Delta_c(j)-2}(2\kappa_c^2 z^2),
\end{equation}
with normalization constant
\begin{equation}
c_n^{(c)}(j)
=
\left[
\frac{2^{\Delta_c(j)+1}\kappa_c^{2\Delta_c(j)}\Gamma(n+1)}
{\Gamma\bigl(n+\Delta_c(j)-1\bigr)}
\right]^{1/2}.
\end{equation}
They satisfy
\begin{equation}
\frac{1}{2\kappa_c^2}
\int_0^{\infty}\dd z\,\sqrt g\,\ee^{-2\Phi_c}|g^{xx}|
\psi_n^{(c)}(j,z)\psi_m^{(c)}(j,z)
=\delta_{nm},
\end{equation}
and the bulk--to--bulk propagator has the spectral representation
\begin{equation}
G_c(j,z,z';t)
=
2\kappa_c^2
\sum_{n=0}^{\infty}
\frac{\psi_n^{(c)}(j,z)\psi_n^{(c)}(j,z')}{-t+m_{c,n}^2(j)},
\label{eq:closed-propagator-spectral}
\end{equation}
with the mass spectrum 
\begin{equation}
\label{eq:Mn-closed-ddvcs}
M_{c,n}^2(j)=8\kappa_c^2\left(n+\frac{\Delta_c(j)}{2}\right)=8\kappa_c^2\left(n+\frac{j+\gamma_c(j)+2}{2}\right).
\end{equation}
The generalized decay constants are defined by the ultraviolet derivative
\begin{equation}
\mathcal F_n^{(c)}(j,\eps)
=
\frac{1}{2\kappa_c^2}\frac{1}{\wt g_5}
\left[-\sqrt g\,\ee^{-2\Phi_c}|g^{xx}|\partial_z\psi_n^{(c)}(j,z)\right]_{z=\eps},
\end{equation}
so that one endpoint of the propagator may be pushed to the boundary as
\begin{equation}
G_c(j,z'\to0,z;t)
\simeq
\frac{\psi_0^{(c)}(j,z'\to0)}{\wt g_5\,\mathcal F_0^{(c)}(j,\eps)}
\sum_{n=0}^{\infty}
\frac{\wt g_5\,\mathcal F_n^{(c)}(j,\eps)\psi_n^{(c)}(j,z)}{-t+m_{c,n}^2(j)}.
\label{eq:near-boundary-factorization-propagator}
\end{equation}
The required $n$-independence follows directly from the universal near-boundary expansion. Since
\begin{equation}
L_n^{\Delta_c(j)-2}(0)=\frac{\Gamma(n+\Delta_c(j)-1)}{\Gamma(\Delta_c(j)-1)\Gamma(n+1)},
\end{equation}
one has
\begin{equation}
\psi_n^{(c)}(j,z'\to0)=c_n^{(c)}(j)L_n^{\Delta_c(j)-2}(0)\,z'^{\Delta_c(j)}[1+\ord(z'^2)],
\end{equation}
while the UV derivative defining $\mathcal F_n^{(c)}$ gives the same factor $c_n^{(c)}(j)L_n^{\Delta_c(j)-2}(0)$ multiplying $\eps^{\Delta_c(j)-4}[1+\ord(\eps^2)]$. Hence this factor cancels in
\begin{equation}
\frac{\psi_n^{(c)}(j,z'\to0)}{\wt g_5\,\mathcal F_n^{(c)}(j,\eps)}
\propto z'^{\Delta_c(j)}\eps^{4-\Delta_c(j)}[1+\ord(z'^2,\eps^2)],
\end{equation}
and the ratio is independent of $n$ to leading near-boundary order; Eq.~\eqref{eq:closed-boundary-mode} fixes the proportionality constant in the present normalization. The same cancellation is the hard-wall statement written with the small-argument Bessel expansion. This is the holographic origin of collinear factorization: the first factor becomes the boundary mode entering the upper three--point function, whereas the second factor becomes the bulk--to--boundary kernel entering the lower hadronic matrix element.

It is therefore natural to define
\begin{equation}
\Psi_j^{(c),\bdry}(z';\eps)=\frac{\psi_0^{(c)}(j,z'\to0)}{\wt g_5\,\mathcal F_0^{(c)}(j,\eps)}
=
-\frac{(\sqrt2\kappa_c z')^{\Delta_c(j)}(\sqrt2\beps_c)^{4-\Delta_c(j)}}{\Delta_c(j)},
\qquad
\beps_c\equiv \kappa_c\eps,
\label{eq:closed-boundary-mode}
\end{equation}
and
\begin{align}
\cH_j^{(c)}(K,z;\eps)
&=\sum_{n=0}^{\infty}
\frac{\wt g_5\,\mathcal F_n^{(c)}(j,\eps)\psi_n^{(c)}(j,z)}{-t+m_{c,n}^2(j)}\nonumber\\
&=\cN_c(j,\bar K_c,\beps_c)
(\sqrt2\kappa_c z)^{\Delta_c(j)}
U\!\left(
\frac{a_{t,c}}{2}+\frac{\Delta_c(j)}{2},
\Delta_c(j)-1;
2\kappa_c^2z^2
\right),
\label{eq:closed-bulk-to-boundary-kernel}
\end{align}
with
\begin{equation}
\cN_c(j,\bar K_c,\beps_c)
=
(\sqrt2\beps_c)^{\Delta_c(j)-4}
\frac{\Gamma\!\left(\frac{a_{t,c}}{2}+\frac{\Delta_c(j)}{2}\right)}{\Gamma\!\left(\Delta_c(j)-2\right)},
\qquad
\bar K_c\equiv \frac{K}{\kappa_c},
\qquad
a_{t,c}\equiv -\frac{t}{4\kappa_c^2}.
\end{equation}
Equation~\eqref{eq:near-boundary-factorization-propagator} is precisely the smooth transition from
Fig.~\ref{TWITTEN-2} to Fig.~\ref{fig:witten-factorized-closed}.
The normalization factor $\cN_c(j,\bar K_c,\beps_c)$ is fixed by the standard normalization condition of bulk-to-boundary propagators in AdS namely $$\cH_j^{(c)}(K,z=\eps;\eps)=1.$$

\subsection{Why the upper integral depends on \texorpdfstring{$\delta_c=j+\Delta_c-2$}{delta=j+Delta-2}}

The fixed--$j$ closed amplitude before evaluating the bulk integrals can be written schematically as
\begin{equation}
\hH^{(c)}(j;s,t,\chi;Q_1^2,Q_2^2)
=
\bar s_c^{\,j}
\widetilde{\cC}_{\gamma\gamma}^{(c)}(j;Q_1^2,Q_2^2,\beps_c)
\widehat d_j(\eta,t)
\cF_N^{(c)}(j;t,\beps_c),
\qquad
\bar s_c\equiv \frac{P\cdot \widetilde{q}}{\kappa_c^2}\simeq \left(\frac{Q^2}{2\kappa_c^2}\right)^j\xi^{-j}.
\label{eq:closed-factorized-schematic}
\end{equation}
The upper vertex is
\begin{equation}
\widetilde{\cC}_{\gamma\gamma}^{(c)}(j;Q_1^2,Q_2^2,\beps_c)
=
\frac{\wt g_5}{2g_5^2}
\int_0^{\infty}\dd z\,\sqrt g\,\ee^{-\Phi_c}
\,z^{4+2(j-2)}
\cK_{\gamma\gamma}^{(c)}(Q_1,Q_2;z)
\,z^{-(j-2)}\Psi_j^{(c),\bdry}(z;\eps),
\label{eq:closed-top-vertex-preconf}
\end{equation}
where
\begin{equation}
\cK_{\gamma\gamma}^{(c)}(Q_1,Q_2;z)=\cV(Q_1,z)\cV(Q_2,z).
\label{eq:closed-TT-photon-kernel}
\end{equation}
Equation~\eqref{eq:closed-TT-photon-kernel} is the TT-projected scalar part of the gauge-invariant photon--photon--spin-$j$ upper vertex, with the scalar projection defined in Eq.~\eqref{eq:TTscalar-invariant}. Explicitly, at leading power the projected tensor structure is summarized by
\begin{equation}
-\frac12 g^T_{\mu\nu}\,\cV^{\mu\nu}_{VV\cO_j}\Big|_{\rm lead}
=\cN_j\,\cV(Q_1,z)\cV(Q_2,z)
+\ord\!\left(\frac{M_N^2}{Q^2},\frac{-t}{Q^2}\right),
\label{eq:app-TT-projected-upper}
\end{equation}
whereas the complementary projections retain the $F_{z\mu}$, $K_0$, and $z\partial_zK_1$ structures. The full Maxwell stress-tensor coupling also contains terms with $F_{z\mu}$ and derivatives of the bulk-to-boundary propagator; in the conformal limit these generate structures proportional to $K_0$ or $z\partial_zK_1$. After the scalar TT projection, such terms feed longitudinal or distinct tensor amplitudes outside the leading-twist unpolarized singlet scalar invariant considered here, as in the TT/LL decompositions of Refs.~\cite{Nishio:2014rya,Nishio:2014eua,Mamo:2021tzd}. The matching claim therefore uses the TT-projected upper vertex, not the full helicity decomposition.
The sector label on the dilaton in Eq.~\eqref{eq:closed-top-vertex-preconf} is a normalization convention for the exchanged closed-channel cubic vertex. In the hard upper-vertex region $z\sim 1/Q$,
\begin{equation}
\ee^{-\Phi_c(z)}-\ee^{-\Phi_o(z)}=-(\kappa_c^2-\kappa_o^2)z^2+\ord(\kappa_X^4z^4),
\end{equation}
so replacing one by the other changes the conformal upper integral only by relative corrections of order $(\kappa_c^2-\kappa_o^2)/Q^2$ plus IR-sensitive normalization effects. It therefore does not affect the leading conformal $C_1$ kernel or the exponent derived below.
In the conformal limit,
\begin{equation}
\cK_{\gamma\gamma}^{(c)}(Q_1,Q_2;z)
\longrightarrow
Q_1Q_2z^2K_1(Q_1z)K_1(Q_2z),
\end{equation}
and the complete $z$--power multiplying the two Bessel functions is
\begin{equation}
(-5)+(4+2(j-2))-(j-2)+\Delta_c(j)+2
=j+\Delta_c(j)-1.
\end{equation}
Thus, after $y=Qz$ with
\begin{equation}
Q_1^2\simeq \left(1+\frac{\eta}{\xi}\right)Q^2,
\qquad
Q_2^2\simeq \left(1-\frac{\eta}{\xi}\right)Q^2,
\end{equation}
the universal upper integral necessarily takes the form
\begin{equation}
\left(1-\frac{\eta^2}{\xi^2}\right)^{1/2}\int_0^{\infty}\dd y\,y^{1+\delta_c(j)}K_1\!\left(y\sqrt{1+\frac{\eta}{\xi}}\right)
K_1\!\left(y\sqrt{1-\frac{\eta}{\xi}}\right),
\end{equation}
with
\begin{equation}
\delta_c(j)=j+\Delta_c(j)-2=2j+\gamma_c(j),
\end{equation}
and the overall kinematic factor coming from the relation $$Q_1Q_2\simeq Q\left(1+\frac{\eta}{\xi}\right)^{1/2}\times Q\left(1-\frac{\eta}{\xi}\right)^{1/2}=Q^2\left(1-\frac{\eta^2}{\xi^2}\right)^{1/2}.$$
The same reasoning goes through verbatim in the even open channel, so the exponent is universally
$\delta_X(j)=j+\Delta_X(j)-2$.

\subsection{Universal upper vertex and the explicit source normalization}

Evaluating \eqref{eq:closed-top-vertex-preconf} with the conformal photon wave functions gives
\begin{equation}
\cC_{\gamma\gamma}^{(c)}(j;Q_1^2,Q_2^2,\beps_c)
=\frac{\wt g_5}{(\beps_c)^{4-\Delta_c(j)}}\times\left(\frac{Q^2}{2\kappa_c^2}\right)^j\times\widetilde{\cC}_{\gamma\gamma}^{(c)}(j;Q_1^2,Q_2^2,\beps_c)=
-\frac{\wt g_5^2}{2g_5^2}
\frac{(\sqrt2)^{4-\Delta_c(j)}}{\Delta_c(j)}
2^j\left(\frac{Q}{2\kappa_c}\right)^{-\delta_c(j)+2j}
C_1\!\left(\delta_c(j),\frac{\eta}{\xi}\right),
\label{eq:closed-top-vertex-afterC1}
\end{equation}
where $C_1$ is defined in \eqref{eq:main-C1-def}.  Using \eqref{eq:main-C1-result}, one obtains
\begin{align}
\cC_{\gamma\gamma}^{(c)}
&=
-\frac{\wt g_5^2}{2g_5^2}
\frac{(\sqrt2)^{4-\Delta_c(j)}}{\Delta_c(j)}
\left(\frac{Q}{2\kappa_c}\right)^{-\delta_c(j)+2j}
2^{\delta_c(j)+j-1}
\frac{\delta_c(j)+2}{\delta_c(j)}
\frac{\Gamma\!\left(\frac{\delta_c(j)}{2}+1\right)^4}{\Gamma\!\left(\delta_c(j)+2\right)}
\nonumber\\
&\hspace{2cm}\times
{}_2F_1\!\left(
\frac{j}{2}+\frac{\gamma_c}{4},
\frac{j+1}{2}+\frac{\gamma_c}{4};
 j+\frac32+\frac{\gamma_c}{2};
 \frac{\eta^2}{\xi^2}
\right).
\label{eq:closed-top-vertex-hypergeom}
\end{align}
Writing
\begin{equation}
\kappa_X=\tilde{\delta}_X\Lambda_{QCD}\quad(X=o,c),
\qquad
\mu_0=\tilde{\delta}_0\Lambda_{QCD},
\qquad
\eps=\frac{\tilde{\delta}_\eps}{\mu},
\end{equation}
one isolates the physical QCD--like scale dependence as
\begin{equation}
\left(\frac{Q}{2\kappa_c}\right)^{-\delta_c(j)+2j}=\left(\frac{Q}{2\kappa_c}\right)^{-\gamma_c(j)}
=
\left(\frac{\mu}{Q}\right)^{\gamma_c(j)}
\left(\frac{\mu_0}{\mu}\right)^{\gamma_c(j)}
\left(\frac{2\tilde{\delta}_c}{\tilde{\delta}_0}\right)^{\gamma_c(j)}.
\label{eq:kappa-to-mu-factorization}
\end{equation}
The remaining $t$-- and $\eta$--independent normalization can be written as a source factor
\begin{align}
\phi_{0}^{c}(j,\tilde{\delta}_c,\tilde{\delta}_0)
&\equiv
-2^{j+\delta_c(j)-2}
\frac{\delta_c(j)+2}{\delta_c(j)\,\Delta_c(j)}
\frac{\Gamma\!\left(\frac{\delta_c(j)}{2}+1\right)^4}{\Gamma\!\left(\delta_c(j)+2\right)}
\left(\sqrt2\right)^{4-\Delta_c(j)}
\left(\frac{2\tilde{\delta}_c}{\tilde{\delta}_0}\right)^{\gamma_c(j)}.
\label{eq:phi0-raw-closed}
\end{align}
This is the explicit leftover source normalization produced by the closed upper vertex in the UV
source convention used here. In this convention the cutoff parameter $\tilde{\delta}_\eps$ has cancelled between the boundary-mode normalization and the source rescaling, so it is not kept as an argument of $\phi_0^X$. At the protected spin--2 point this normalization is finite:
\begin{equation}
\gamma_c(2)=0,
\qquad
\delta_c(2)=\Delta_c(2)=4,
\qquad
\phi_0^c(2,\tilde\delta_c,\tilde\delta_0)=-\frac45,
\label{eq:phi0-closed-j2-check}
\end{equation}
before the overall factor $\widetilde g_5^2/g_5^2$ is included. On the QCD side, in the LO forward normalization convention of Appendix~\ref{app:pqcd},
\begin{equation}
c_2^{-,(0)}=-\frac{\gamma_2^{\Sigma g,(0)}}{\gamma_2^{+,(0)}-\gamma_2^{-,(0)}}
=\frac{T_F n_f}{2C_F+T_F n_f},
\label{eq:cminus-j2-finite-check}
\end{equation}
so the source-factor/forward-normalization matching is finite at the structural anchor. The minus sign in Eq.~\eqref{eq:phi0-closed-j2-check} is not an invariant obstruction: the phase of the spin-$j$ source/eigenoperator and the sign convention for the matched moment can be reversed, while the physical product in Eq.~\eqref{eq:main-moment-dictionary} is unchanged. The lower soft-wall moment is also finite there for the twist values used in this model.

The crucial point is that none of the soft--wall input used for the nucleon enters
\eqref{eq:closed-top-vertex-hypergeom}.  The entire upper vertex is fixed by pure $\AdS$ photon
wave functions and by the quantum numbers of the exchanged spin--$j$ field.  This is the precise
sense in which the upper vertex is universal and model independent.

\subsection{Model--dependent lower vertex in the soft--wall model}

All model dependence is isolated in the lower impact factor
\begin{equation}
\cF_N^{(c)}(j;t,\beps_c)
=
\frac{\wt g_5}{2}
\int_0^{\infty}\dd z\,\sqrt g\,\ee^{-\Phi_c}
\,z^{1+2(j-2)}
\bigl(\psi_R^2+\psi_L^2\bigr)
\,z^{-(j-2)}\cH_j^{(c)}(K,z;\eps).
\end{equation}
With $u=2\kappa_c^2z^2$ and the standard integral
\begin{equation}
\int_0^{\infty}\dd u\,u^{\lambda-1}\ee^{-u}U(a,b;u)
=
\frac{\Gamma(\lambda)\Gamma(\lambda-b+1)}{\Gamma(a+\lambda-b+1)},
\end{equation}
the lower vertex reduces to pure gamma functions.  Separating the right-- and left--handed nucleon
contributions gives
\begin{subequations}
\label{eq:chatF-closed-main-updated}
\begin{align}
\widehat{\cF}_{R}^{(c)}(j;t)=2\times\frac{(\beps_c)^{4-\Delta_c(j)}}{\wt g_5}\times\cF_N^{(c)}(j;t,\beps_c)
&=
(\tau-1)
\frac{
\Gamma\!\left(\tau-1-\frac{\gamma_c}{2}\right)
\Gamma\!\left(j+\tau-1+\frac{\gamma_c}{2}\right)
\Gamma\!\left(a_{t,c}+j+\gamma_c\right)
}{
\Gamma\!\left(j+\gamma_c\right)
\Gamma(\tau)
\Gamma\!\left(a_{t,c}+j+\tau-1+\frac{\gamma_c}{2}\right)
},
\\[1ex]
\widehat{\cF}_{L}^{(c)}(j;t)
=2\times\frac{(\beps_c)^{4-\Delta_c(j)}}{\wt g_5}\times\cF_N^{(c)}(j;t,\beps_c)&=
\frac{
\Gamma\!\left(\tau-\frac{\gamma_c}{2}\right)
\Gamma\!\left(j+\tau+\frac{\gamma_c}{2}\right)
\Gamma\!\left(a_{t,c}+j+\gamma_c\right)
}{
\Gamma\!\left(j+\gamma_c\right)
\Gamma(\tau)
\Gamma\!\left(a_{t,c}+j+\tau+\frac{\gamma_c}{2}\right)
},
\\[1ex]
\widehat{\cF}_{N}^{(c)}(j;t)
&=
\frac12\Big[
\widehat{\cF}_{R}^{(c)}(j;t)+
\widehat{\cF}_{L}^{(c)}(j;t)
\Big],
\end{align}
\end{subequations}
with $a_{t,c}=-t/(4\kappa_c^2)$.  Equations~\eqref{eq:chatF-closed-main-updated} make the
factorization transparent: the upper vertex is universal, the lower vertex carries all model
information, and in the present paper that information is taken from the soft--wall model.

Combining the upper and lower pieces yields the closed fixed--$j$ amplitude in the form quoted in
Eq.~\eqref{eq:main-holo-fixedj}, namely
\begin{equation}
\hH^{(c)}(j)
=
\xi^{-j}
\left(\frac{\mu}{Q}\right)^{\gamma_c}
{}_2F_1\!\left(
\frac{j}{2}+\frac{\gamma_c}{4},
\frac{j+1}{2}+\frac{\gamma_c}{4};
 j+\frac32+\frac{\gamma_c}{2};
 \frac{\eta^2}{\xi^2}
\right)
\left(\frac{\mu_0}{\mu}\right)^{\gamma_c}
\phi_0^c(j,\tilde{\delta}_c,\tilde{\delta}_0)
\widehat d_j(\eta,t)
\left(\frac{\wt g_5^2}{g_5^2}\right)
\widehat{\cF}_N^{(c)}(j;t).
\label{eq:closed-final-main-text}
\end{equation}


\subsection{Even--spin open channel as the parallel Nishio--Watari construction}
\label{sec:open-channel}

The even open channel is not derived from the closed channel by fiat. The Nishio--Watari open-string construction supplies the parallel fixed--$j$ upper vertex, whose projected scalar part is the object relevant for the unpolarized singlet vector invariant amplitude. In the conventions used here this upper vertex is
\begin{equation}
\widetilde{\cC}_{\gamma\gamma}^{(o)}(j;Q_1^2,Q_2^2,\beps_o)
=
\frac{g_5}{2g_5^2}
\int_0^{\infty}\dd z\,\sqrt g\,\ee^{-\Phi_o}
\,z^{4+2(j-2)}
\cK_{\gamma\gamma}^{(o)}(Q_1,Q_2;z)
\,z^{-(j-2)}\Psi_j^{(o),\bdry}(z;\eps),
\label{eq:app-open-top-vertex-preconf}
\end{equation}
with
\begin{subequations}
\label{eq:app-open-kernel-boundary}
\begin{align}
\cK_{\gamma\gamma}^{(o)}(Q_1,Q_2;z)&=\cV(Q_1,z)\cV(Q_2,z),
\\
\Psi_j^{(o),\bdry}(z;\eps)
&=
-\frac{(\sqrt2\kappa_o z)^{\Delta_o(j)}(\sqrt2\beps_o)^{4-\Delta_o(j)}}{\Delta_o(j)}.
\end{align}
\end{subequations}
Taking the same conformal photon limit $\cV(Q_i,z)\to Q_i zK_1(Q_i z)$ gives the open-channel power count
\begin{equation}
 z^{-5}\,z^{4+2(j-2)}\,z^{-(j-2)}\,z^{\Delta_o(j)}\,z^2
 =z^{j+\Delta_o(j)-1}=z^{1+\delta_o(j)},
\label{eq:app-open-z-count}
\end{equation}
so the open upper vertex contains the same Bessel product integral
\begin{equation}
C_1\!\left(\delta_o(j),\frac{\eta}{\xi}\right)
=
\left(1-\frac{\eta^2}{\xi^2}\right)^{1/2}
\int_0^\infty\dd y\,y^{1+\delta_o(j)}
K_1\!\left(y\sqrt{1+\frac{\eta}{\xi}}\right)
K_1\!\left(y\sqrt{1-\frac{\eta}{\xi}}\right),
\label{eq:app-open-C1}
\end{equation}
and hence the same Gauss-hypergeometric kernel with $\gamma_c\to\gamma_o$. This is precisely the $C_1$ integral and hypergeometric transformation displayed by Nishio--Watari in Appendix~B of Ref.~\cite{Nishio:2014eua}. The open field $h^o_{\mu\nu}$ should therefore be read here as the effective even-spin open Reggeon/tensor exchange after projection onto the same scalar invariant, not as a claim about the full tensor structure of the open-string sector.

Only after the open and closed projected vertices have both been specified do we summarize their relation by the direct replacement rule \eqref{eq:main-replacement-rule}. The rule is therefore a shorthand between two fixed--$j$ Witten vertices, not a derivation of the open sector from the closed one. It acts on the projected upper scalar kernel and the resulting fixed--$j$ invariant amplitude, not on the full open-string tensor sector.  The open trajectory is
\begin{equation}
\Delta_o(j)=2+j+\gamma_o(j)=2+\sqrt{\sqrt\lambda\,(j-j_{0o})},
\qquad
j_{0o}=1-\frac{1}{\sqrt\lambda},
\end{equation}
so that
\begin{equation}
\delta_o(j)=j+\Delta_o(j)-2=2j+\gamma_o(j).
\end{equation}
Because the upper vertex depends only on the conformal photon wave functions and on the spin and
dimension of the exchanged field, it has the same functional form as in the closed channel,
with the substitutions $\Delta_c\to \Delta_o$ and $\gamma_c\to \gamma_o$.  Thus the open source
factor is obtained directly from \eqref{eq:phi0-raw-closed} by replacement,
\begin{equation}
\phi_0^{o}(j,\tilde{\delta}_o,\tilde{\delta}_0)
=\phi_0^{c}(j,\tilde{\delta}_c,\tilde{\delta}_0)
\Big|_{\delta_c\to\delta_o,\,\Delta_c\to\Delta_o,\,\gamma_c\to\gamma_o}.
\label{eq:phi0-open-from-closed}
\end{equation}
The lower soft--wall moments are likewise obtained by the channel replacement
\begin{subequations}
\label{eq:chatF-open-main-updated}
\begin{align}
\widehat{\cF}_{R}^{(o)}(j;t)
&=
(\tau-1)
\frac{
\Gamma\!\left(\tau-1-\frac{\gamma_o}{2}\right)
\Gamma\!\left(j+\tau-1+\frac{\gamma_o}{2}\right)
\Gamma\!\left(a_{t,o}+j+\gamma_o\right)
}{
\Gamma\!\left(j+\gamma_o\right)
\Gamma(\tau)
\Gamma\!\left(a_{t,o}+j+\tau-1+\frac{\gamma_o}{2}\right)
},
\\[1ex]
\widehat{\cF}_{L}^{(o)}(j;t)
&=
\frac{
\Gamma\!\left(\tau-\frac{\gamma_o}{2}\right)
\Gamma\!\left(j+\tau+\frac{\gamma_o}{2}\right)
\Gamma\!\left(a_{t,o}+j+\gamma_o\right)
}{
\Gamma\!\left(j+\gamma_o\right)
\Gamma(\tau)
\Gamma\!\left(a_{t,o}+j+\tau+\frac{\gamma_o}{2}\right)
},
\\[1ex]
\widehat{\cF}_{N}^{(o)}(j;t)
&=
\frac12\Big[
\widehat{\cF}_{R}^{(o)}(j;t)+
\widehat{\cF}_{L}^{(o)}(j;t)
\Big],
\end{align}
\end{subequations}
with $a_{t,o}=-t/(4\kappa_o^2)$. For the Sommerfeld--Watson reconstruction the gamma-function expressions in Eqs.~\eqref{eq:chatF-closed-main-updated} and \eqref{eq:chatF-open-main-updated} are understood by analytic continuation from the physical even-spin sequence. The contours $\cC_X$ are chosen to avoid isolated poles generated by the soft-wall lower vertex; such poles are model-dependent lower-moment singularities, not additional ultraviolet Wilson kernels. In applications they are treated by the usual contour-deformation or residue prescription. The final open fixed--$j$ amplitude is therefore
\begin{equation}
\hH^{(o)}(j)
=
\xi^{-j}
\left(\frac{\mu}{Q}\right)^{\gamma_o}
{}_2F_1\!\left(
\frac{j}{2}+\frac{\gamma_o}{4},
\frac{j+1}{2}+\frac{\gamma_o}{4};
 j+\frac32+\frac{\gamma_o}{2};
 \frac{\eta^2}{\xi^2}
\right)
\left(\frac{\mu_0}{\mu}\right)^{\gamma_o}
\phi_0^o(j,\tilde{\delta}_o,\tilde{\delta}_0)
\widehat d_j(\eta,t)
\widehat{\cF}_N^{(o)}(j;t).
\label{eq:open-final-main-text}
\end{equation}
This is a notation-saving summary after the parallel Nishio--Watari open Witten vertex has been specified; it is not a derivation of the open sector from the closed replacement rule.


\subsection{Reconstruction of the physical Compton form factor and operator interpretation}
\label{sec:reconstruction}

The fixed--$j$ amplitudes \eqref{eq:closed-final-main-text} and \eqref{eq:open-final-main-text}
are the building blocks of the physical \DDVCS\ Compton form factor.  The reggeized amplitude is
obtained by the even--signature Sommerfeld--Watson transform,
\begin{equation}
\hH(\bar s,\bar t,\chi;Q_1^2,Q_2^2)
=
\sum_{X=o,c}
\int_{\cC_X}\frac{\dd j}{2\pi\ii}\,
\xi_+(j)\,\hH^{(X)}(j),
\qquad
\xi_+(j)=\frac{1+\ee^{-\ii\pi j}}{\sin\pi j},
\label{eq:SW-reconstruction}
\end{equation}
with only even spins contributing in the present sector.

After holographic collinear factorization the operator interpretation is immediate.  The upper
vertex is the universal three--point function of two electromagnetic currents and a normalized
spin--$j$ operator.  The lower vertex is the matrix element of that same operator between
nucleon states.  In symbolic form,
\begin{equation}
\text{upper vertex}
\sim \langle J^\mu J^\nu \widetilde{\cO}^{(j)}_{X}\rangle,
\qquad
\text{lower vertex}
\sim \langle P'|\cO_{X}^{(j)}|P\rangle.
\end{equation}
The universal hypergeometric function is therefore the holographic analogue of the perturbative
Wilson coefficient, whereas the bottom moments $\widehat{\cF}_N^{(X)}$ are the holographic analogue
of the nonperturbative conformal moments.

This factorized structure also makes the model dependence completely explicit.  The universal upper
vertex does not know about the soft-wall.  The soft--wall model enters only through the lower
hadronic integral, and hence only through the bottom moments
$\widehat{\cF}_N^{(X)}(j;t)$.  Any other IR model would modify those moments while leaving the
hypergeometric hard kernel unchanged.


\subsection{Useful algebra for the lower soft--wall moments}
\label{app:bottom-details}

For completeness we record the lower--vertex integral in a form that makes the gamma--function
reduction completely transparent.  Inserting \eqref{eq:closed-bulk-to-boundary-kernel} and the
soft--wall nucleon wave functions into the lower vertex gives integrals of the schematic form
\begin{equation}
\int_0^{\infty}\dd z\,z^{\alpha-1}\ee^{-2\kappa_X^2z^2}
U\!\left(\frac{a_{t,X}}{2}+\frac{\Delta_X}{2},\Delta_X-1;2\kappa_X^2z^2\right),
\end{equation}
which, after $u=2\kappa_X^2z^2$, reduce to
\begin{equation}
\int_0^{\infty}\dd u\,u^{\lambda-1}\ee^{-u}U(a,b;u)
=
\frac{\Gamma(\lambda)\Gamma(\lambda-b+1)}{\Gamma(a+\lambda-b+1)}.
\end{equation}
This is the elementary identity behind the gamma--function expressions
\eqref{eq:chatF-closed-main-updated} and \eqref{eq:chatF-open-main-updated}.


\subsection{Unified fixed--\texorpdfstring{$j$}{j} formula}
\label{app:unified-formula}

It is sometimes useful to write the open and closed results in a single line.  Defining
\begin{equation}
\mathfrak g_o\equiv 1,
\qquad
\mathfrak g_c\equiv \frac{\wt g_5^2}{g_5^2},
\end{equation}
we may combine the two channels as
\begin{align}
\hH^{(X)}(j)
&=
\xi^{-j}
\left(\frac{\mu}{Q}\right)^{\gamma_X}
{}_2F_1\!\left(
\frac{j}{2}+\frac{\gamma_X}{4},
\frac{j+1}{2}+\frac{\gamma_X}{4};
 j+\frac32+\frac{\gamma_X}{2};
 \frac{\eta^2}{\xi^2}
\right)
\left(\frac{\mu_0}{\mu}\right)^{\gamma_X}\nonumber\\
&\quad\times
\phi_0^X(j,\tilde{\delta}_X,\tilde{\delta}_0)
\widehat d_j(\eta,t)
\mathfrak g_X\widehat{\cF}_N^{(X)}(j;t),
\qquad X=o,c.
\end{align}
This compact form makes the universal structure obvious: channel dependence enters only through the
trajectory data $(\Delta_X,\gamma_X)$, the overall coupling factor $\mathfrak g_X$, and the lower
moments $\widehat{\cF}_N^{(X)}$.


\section{Detailed pQCD conformal--OPE derivation}
\label{app:pqcd}
\label{app:pqcd-singlet-H}

In this appendix we summarize the twist--two pQCD result for the unpolarized singlet vector
Compton form factor in the conformal basis that diagonalizes the LO singlet evolution.  We keep
only the singlet sector and work throughout in the $\pm$ basis.  To match the convention used in
the main text, we replace the conformal-spin label of the conformal-OPE reference \cite{Kumericki:2007sa} by
$j_{\rm ref}\to j-1$.  With this shift the vector projector
$[1-(-1)^{j_{\rm ref}}]$ becomes $[1+(-1)^j]$, and the factor $\xi^{-j_{\rm ref}-1}$ becomes
$\xi^{-j}$.

\subsection{Scheme and sign conventions}
\label{app:pqcd-scheme-sign}

Our anomalous dimensions are defined by
\begin{equation}
\mu\frac{\dd}{\dd\mu}H_j^a(\mu)=-\gamma_j^a(\alpha_s)H_j^a(\mu),
\qquad a=\pm,
\label{eq:app-pqcd-sign-convention}
\end{equation}
so that positive $\gamma_j^a$ suppresses the corresponding conformal moment when evolved to larger scales. The shift $j_{\rm ref}=j-1$ is only a relabeling from the convention of Ref.~\cite{Kumericki:2007sa} to the physical spin-$j$ convention used in the holographic fixed-$j$ amplitude. Wilson coefficients are treated as row vectors and conformal moments as column vectors; this fixes the placement of $U_j$ and $U_j^{-1}$ below. Finally, the finite constants $c_j^\pm$ are normalization-dependent forward Wilson-coefficient entries. A rescaling of the $\pm$ eigenvectors rescales $c_j^\pm$ and $H_j^\pm$ inversely and leaves the products $C_j^\pm H_j^\pm$ unchanged.

\paragraph{CS representation and beta-proportional terms.}
The kernel-level statement below uses the conformal partial-wave/CS representation. In that representation beta-proportional conformal-anomaly terms and finite scheme transformations belong to the coefficient/evolution bookkeeping that relates scales and schemes. At the single matching point $Q=\mu=\mu_0=\mu_\ast$ they do not define a second $\eta/\xi$ kernel for the projected invariant; they modify $\gamma_j^\pm$, $c_j^\pm$, and evolution factors. The matching is therefore not a scheme-independent claim about complete NNLO coefficient functions in arbitrary form, but a fixed-scale statement about the conformal kernel and matched moments.

It is convenient to define the fixed--$j$ singlet contribution by
\begin{equation}
\widehat{\mathcal H}^{\rm sing}_{\rm pQCD}(j;\xi,\eta,t,Q^2;\mu)
\equiv
\xi^{-j}\,C_j^+\!\left(\frac{\eta}{\xi},\frac{Q^2}{\mu^2};\alpha_s(\mu)\right)
H_j^+(\eta,t;\mu)
+
\xi^{-j}\,C_j^-\!\left(\frac{\eta}{\xi},\frac{Q^2}{\mu^2};\alpha_s(\mu)\right)
H_j^-(\eta,t;\mu).
\label{eq:app-pqcd-fixedj-def}
\end{equation}
The singlet part of the vector CFF is then
\begin{equation}
\mathcal H_{\rm pQCD}^{\rm sing}(\xi,\eta,t,Q^2;\mu)
=
\sum_{j=1}^{\infty}
[1+(-1)^j]\,
\widehat{\mathcal H}^{\rm sing}_{\rm pQCD}(j;\xi,\eta,t,Q^2;\mu),
\label{eq:app-pqcd-H-singlet-sum}
\end{equation}
up to the overall singlet charge factor $Q_S^2$, which we suppress for notational simplicity.
The projector $[1+(-1)^j]$ enforces even $j$, so the first nonvanishing term is $j=2$.

\subsection{LO diagonalization of the singlet sector}

Let $H_j^\Sigma$ and $H_j^g$ denote the conformal moments of the singlet quark and gluon GPDs
in the unpolarized vector sector.  The LO anomalous-dimension matrix in our $j$-convention is
\begin{equation}
\gamma_j^{(0)}
=
\begin{pmatrix}
\gamma_j^{\Sigma\Sigma,(0)} & \gamma_j^{\Sigma g,(0)} \\
\gamma_j^{g\Sigma,(0)} & \gamma_j^{gg,(0)}
\end{pmatrix},
\end{equation}
with
\begin{subequations}
\begin{align}
\gamma_j^{\Sigma\Sigma,(0)}
&=
-C_F\left[
3+\frac{2}{j(j+1)}-4S_1(j)
\right],
\\
\gamma_j^{\Sigma g,(0)}
&=
-4n_fT_F\,
\frac{j^2+j+2}{j(j+1)(j+2)},
\\
\gamma_j^{g\Sigma,(0)}
&=
-2C_F\,
\frac{j^2+j+2}{(j-1)j(j+1)},
\\
\gamma_j^{gg,(0)}
&=
C_A\left[
4S_1(j)+\frac{4}{j(j+1)}-\frac{12}{(j-1)(j+2)}
\right]-\beta_0.
\end{align}
\end{subequations}
Here
\begin{equation}
S_1(j)\equiv \psi(j+1)+\gamma_E,
\qquad
C_F=\frac{N_c^2-1}{2N_c},
\qquad
C_A=N_c,
\qquad
T_F=\frac12,
\qquad
\beta_0=\frac{11}{3}C_A-\frac{4}{3}T_F n_f.
\end{equation}

The two LO eigen-anomalous dimensions are
\begin{equation}
\gamma_j^{\pm,(0)}
=
\frac12\left[
\gamma_j^{\Sigma\Sigma,(0)}+\gamma_j^{gg,(0)}
\pm
\sqrt{
\left(\gamma_j^{\Sigma\Sigma,(0)}-\gamma_j^{gg,(0)}\right)^2
+4\,\gamma_j^{\Sigma g,(0)}\gamma_j^{g\Sigma,(0)}
}
\right].
\label{eq:app-pqcd-gamma-pm}
\end{equation}

We define
\begin{equation}
a_j\equiv
\frac{\gamma_j^{gg,(0)}-\gamma_j^{-,(0)}}{\gamma_j^{g\Sigma,(0)}},
\qquad
b_j\equiv
\frac{\gamma_j^{\Sigma\Sigma,(0)}-\gamma_j^{+,(0)}}{\gamma_j^{\Sigma g,(0)}},
\qquad
D_j\equiv 1-a_j b_j .
\label{eq:app-pqcd-abD}
\end{equation}
Then the $\pm$ basis is introduced by
\begin{equation}
\begin{pmatrix}
H_j^+(\eta,t;\mu_0)\\[0.5ex]
H_j^-(\eta,t;\mu_0)
\end{pmatrix}
=
U_j
\begin{pmatrix}
H_j^\Sigma(\eta,t;\mu_0)\\[0.5ex]
H_j^g(\eta,t;\mu_0)
\end{pmatrix},
\qquad
U_j\equiv
\begin{pmatrix}
1 & a_j\\[1ex]
b_j & 1
\end{pmatrix}.
\label{eq:app-pqcd-U-def}
\end{equation}
Explicitly,
\begin{subequations}
\begin{align}
H_j^+(\eta,t;\mu_0)
&=
H_j^\Sigma(\eta,t;\mu_0)+a_j\,H_j^g(\eta,t;\mu_0),
\\
H_j^-(\eta,t;\mu_0)
&=
b_j\,H_j^\Sigma(\eta,t;\mu_0)+H_j^g(\eta,t;\mu_0).
\end{align}
\end{subequations}
The inverse relation is
\begin{equation}
U_j^{-1}
=
\frac{1}{D_j}
\begin{pmatrix}
1 & -a_j\\[1ex]
-b_j & 1
\end{pmatrix},
\end{equation}
so that
\begin{subequations}
\label{eq:app-pqcd-HSigmag-inverse}
\begin{align}
H_j^\Sigma(\eta,t;\mu_0)
&=
\frac{H_j^+(\eta,t;\mu_0)-a_j H_j^-(\eta,t;\mu_0)}{D_j},
\\[1ex]
H_j^g(\eta,t;\mu_0)
&=
\frac{-b_j H_j^+(\eta,t;\mu_0)+H_j^-(\eta,t;\mu_0)}{D_j}.
\end{align}
\end{subequations}

The normalization of the $\pm$ eigenvectors is not unique.  Any rescaling
\begin{equation}
H_j^\pm \;\longrightarrow\; N_j^\pm\,H_j^\pm,
\qquad
C_j^\pm \;\longrightarrow\; (N_j^\pm)^{-1} C_j^\pm,
\label{eq:app-pqcd-normalization-freedom}
\end{equation}
leaves the physical Compton form factor unchanged.  The convention in
\eqref{eq:app-pqcd-U-def} is the convenient one in which the transformation matrix has unit
diagonal entries.

\subsection{Conformal fixed point}

At a conformal fixed point $\alpha_s=\alpha_s^\ast$, the LO singlet eigenchannels evolve
autonomously,
\begin{equation}
\mu\frac{\dd}{\dd\mu}H_j^\pm(\eta,t;\mu)
=
-\gamma_j^\pm(\alpha_s^\ast)\,
H_j^\pm(\eta,t;\mu),
\qquad
\gamma_j^\pm(\alpha_s^\ast)
=
\frac{\alpha_s^\ast}{2\pi}\gamma_j^{\pm,(0)}+\ord\!\big((\alpha_s^\ast)^2\big),
\end{equation}
so that
\begin{equation}
H_j^\pm(\eta,t;\mu)
=
\left(\frac{\mu_0}{\mu}\right)^{\gamma_j^\pm(\alpha_s^\ast)}
H_j^\pm(\eta,t;\mu_0).
\label{eq:app-pqcd-fixed-point-evolution}
\end{equation}

In the same conformal limit, the Wilson coefficients are fixed by conformal symmetry up to the
forward DIS normalization constants $c_j^\pm(\alpha_s^\ast)$.  After the replacement
$j_{\rm ref}\to j-1$, the vector coefficients take the form
\begin{equation}
C_j^\pm\!\left(\frac{\eta}{\xi},\frac{Q^2}{\mu^2};\alpha_s^\ast\right)
=
c_j^\pm(\alpha_s^\ast)
\left(\frac{\mu}{Q}\right)^{\gamma_j^\pm(\alpha_s^\ast)}
{}_2F_1\!\left(
\frac{j}{2}+\frac{\gamma_j^\pm(\alpha_s^\ast)}{4},
\frac{j+1}{2}+\frac{\gamma_j^\pm(\alpha_s^\ast)}{4};
j+\frac32+\frac{\gamma_j^\pm(\alpha_s^\ast)}{2};
\frac{\eta^2}{\xi^2}
\right).
\label{eq:app-pqcd-fixed-point-Wilson}
\end{equation}
Therefore
\begin{align}
&\widehat{\mathcal H}^{\rm sing}_{\rm pQCD}(j;\xi,\eta,t,Q^2;\mu)
=\sum_{a=\pm}
\xi^{-j}\,c_j^a(\alpha_s^\ast)
\left(\frac{\mu}{Q}\right)^{\gamma_j^a(\alpha_s^\ast)}
\cK_j^{(\gamma_j^a(\alpha_s^\ast))}(\vartheta)
H_j^a(\eta,t;\mu)
\nonumber\\
&=\sum_{a=\pm}
\xi^{-j}\,c_j^a(\alpha_s^\ast)
\left(\frac{\mu}{Q}\right)^{\gamma_j^a(\alpha_s^\ast)}
\cK_j^{(\gamma_j^a(\alpha_s^\ast))}(\vartheta)
\left(\frac{\mu_0}{\mu}\right)^{\gamma_j^a(\alpha_s^\ast)}
H_j^a(\eta,t;\mu_0).
\label{eq:app-pqcd-fixed-point-fixedj}
\end{align}
This is the fixed-point pQCD analogue of the fixed--$j$ holographic amplitude in the main text.

\subsection{LO running-coupling modification}

In real QCD the coupling runs.  At LO,
\begin{equation}
\mu\frac{\dd\alpha_s}{\dd\mu}
=
-\frac{\beta_0}{2\pi}\alpha_s^2
+\ord(\alpha_s^3).
\label{eq:app-pqcd-alpha-running}
\end{equation}
The conformal moments still evolve diagonally in the $\pm$ basis at LO,
\begin{equation}
\mu\frac{\dd}{\dd\mu}H_j^\pm(\eta,t;\mu)
=
-\frac{\alpha_s(\mu)}{2\pi}\gamma_j^{\pm,(0)}\,H_j^\pm(\eta,t;\mu),
\end{equation}
with solution
\begin{equation}
H_j^\pm(\eta,t;\mu)
=
\mathcal E_j^\pm(\mu,\mu_0)\,H_j^\pm(\eta,t;\mu_0),
\qquad
\mathcal E_j^\pm(\mu,\mu_0)
=
\left[
\frac{\alpha_s(\mu)}{\alpha_s(\mu_0)}
\right]^{\gamma_j^{\pm,(0)}/\beta_0}.
\label{eq:app-pqcd-LO-evolution}
\end{equation}
Hence the pure power law \((\mu_0/\mu)^{\gamma_j^\pm(\alpha_s^\ast)}\) of the fixed-point theory
is replaced by the usual logarithmic scaling violation controlled by the running coupling.

At the same level of approximation, the conformal Wilson coefficient can be RG-improved as
\begin{align}
C_j^\pm\!\left(\frac{\eta}{\xi},\frac{Q^2}{\mu^2};\alpha_s\right)
&\simeq
c_j^\pm(\alpha_s(Q))
{}_2F_1\!\left(
\frac{j}{2}+\frac{\gamma_j^\pm(\alpha_s(Q))}{4},
\frac{j+1}{2}+\frac{\gamma_j^\pm(\alpha_s(Q))}{4};
j+\frac32+\frac{\gamma_j^\pm(\alpha_s(Q))}{2};
\frac{\eta^2}{\xi^2}
\right)
\left[
\frac{\alpha_s(Q)}{\alpha_s(\mu)}
\right]^{\gamma_j^{\pm,(0)}/\beta_0},
\label{eq:app-pqcd-LO-running-Wilson}
\end{align}
where at LO one may take
\begin{equation}
\gamma_j^\pm(\alpha_s(Q))
=
\frac{\alpha_s(Q)}{2\pi}\,\gamma_j^{\pm,(0)}.
\end{equation}
Combining \eqref{eq:app-pqcd-LO-evolution} and \eqref{eq:app-pqcd-LO-running-Wilson}, one finds
\begin{align}
&\widehat{\mathcal H}^{\rm sing}_{\rm pQCD}(j;\xi,\eta,t,Q^2;\mu)
=\sum_{a=\pm}
\xi^{-j}\,c_j^a(\alpha_s(Q))
\cK_j^{(\gamma_j^a(\alpha_s(Q)))}(\vartheta)
\left[
\frac{\alpha_s(Q)}{\alpha_s(\mu_0)}
\right]^{\gamma_j^{a,(0)}/\beta_0}
H_j^a(\eta,t;\mu_0).
\label{eq:app-pqcd-final-LO-running}
\end{align}
Thus LO running preserves the same factorized logic as the conformal fixed-point formula, but the
pure power-law scaling in \(\mu\) is replaced by the usual logarithmic QCD evolution.

\subsection{Singlet vector Compton form factor in the quark--gluon basis}
\label{app:pqcd-singlet-H-SigmaG}

The $\pm$ basis is convenient because it diagonalizes the LO singlet evolution.  However, the
same fixed--$j$ pQCD amplitude can be written directly in the quark--gluon basis
$\{\Sigma,g\}$.  Since the conformal moments form a column vector while the Wilson coefficients
form a row vector, the two basis changes are
\begin{equation}
\begin{pmatrix}
H_j^+(\eta,t;\mu)\\[0.5ex]
H_j^-(\eta,t;\mu)
\end{pmatrix}
=
U_j
\begin{pmatrix}
H_j^\Sigma(\eta,t;\mu)\\[0.5ex]
H_j^g(\eta,t;\mu)
\end{pmatrix},
\qquad
\bigl(
C_j^\Sigma,C_j^g
\bigr)
=
\bigl(
C_j^+,C_j^-
\bigr)\,U_j.
\label{eq:app-pqcd-row-column-transform}
\end{equation}
Thus
\begin{subequations}
\label{eq:app-pqcd-CSigma-Cg-from-Cpm}
\begin{align}
C_j^\Sigma\!\left(\frac{\eta}{\xi},\frac{Q^2}{\mu^2};\alpha_s(\mu)\right)
&=
C_j^+\!\left(\frac{\eta}{\xi},\frac{Q^2}{\mu^2};\alpha_s(\mu)\right)
+
b_j\,
C_j^-\!\left(\frac{\eta}{\xi},\frac{Q^2}{\mu^2};\alpha_s(\mu)\right),
\\[1ex]
C_j^g\!\left(\frac{\eta}{\xi},\frac{Q^2}{\mu^2};\alpha_s(\mu)\right)
&=
a_j\,
C_j^+\!\left(\frac{\eta}{\xi},\frac{Q^2}{\mu^2};\alpha_s(\mu)\right)
+
C_j^-\!\left(\frac{\eta}{\xi},\frac{Q^2}{\mu^2};\alpha_s(\mu)\right).
\end{align}
\end{subequations}
Therefore the fixed--$j$ singlet amplitude can be written equivalently as
\begin{equation}
\widehat{\mathcal H}^{\rm sing}_{\rm pQCD}(j;\xi,\eta,t,Q^2;\mu)
=
\xi^{-j}\,C_j^\Sigma\!\left(\frac{\eta}{\xi},\frac{Q^2}{\mu^2};\alpha_s(\mu)\right)
H_j^\Sigma(\eta,t;\mu)
+
\xi^{-j}\,C_j^g\!\left(\frac{\eta}{\xi},\frac{Q^2}{\mu^2};\alpha_s(\mu)\right)
H_j^g(\eta,t;\mu).
\label{eq:app-pqcd-Hj-Sigma-g}
\end{equation}
Hence the singlet vector Compton form factor in the quark--gluon basis is
\begin{equation}
\mathcal H_{\rm pQCD}^{\rm sing}(\xi,\eta,t,Q^2;\mu)
=
\sum_{j=1}^{\infty}
[1+(-1)^j]\,
\xi^{-j}
\left[
C_j^\Sigma\,H_j^\Sigma
+
C_j^g\,H_j^g
\right].
\label{eq:app-pqcd-H-Sigma-g-full}
\end{equation}

If one wishes to express the result directly in terms of the quark--gluon input moments at
$\mu_0$, one may first evolve in the diagonal basis and only then rotate back.  Defining
\begin{equation}
E_j^{(\pm)}(\mu,\mu_0)
=
\begin{pmatrix}
\mathcal E_j^+(\mu,\mu_0) & 0\\[0.5ex]
0 & \mathcal E_j^-(\mu,\mu_0)
\end{pmatrix},
\end{equation}
the effective quark--gluon-basis coefficient row vector is
\begin{equation}
\bigl(
C_{j,\rm eff}^{\Sigma},C_{j,\rm eff}^{g}
\bigr)
=
\bigl(
C_j^+\,\mathcal E_j^+,
C_j^-\,\mathcal E_j^-
\bigr)\,U_j.
\label{eq:app-pqcd-Ceff-sigg-def}
\end{equation}
Explicitly,
\begin{subequations}
\label{eq:app-pqcd-Ceff-sigg-components}
\begin{align}
C_{j,\rm eff}^{\Sigma}
&=
\mathcal E_j^+(\mu,\mu_0)\,
C_j^+\!\left(\frac{\eta}{\xi},\frac{Q^2}{\mu^2};\alpha_s(\mu)\right)
+
b_j\,\mathcal E_j^-(\mu,\mu_0)\,
C_j^-\!\left(\frac{\eta}{\xi},\frac{Q^2}{\mu^2};\alpha_s(\mu)\right),
\\[1ex]
C_{j,\rm eff}^{g}
&=
a_j\,\mathcal E_j^+(\mu,\mu_0)\,
C_j^+\!\left(\frac{\eta}{\xi},\frac{Q^2}{\mu^2};\alpha_s(\mu)\right)
+
\mathcal E_j^-(\mu,\mu_0)\,
C_j^-\!\left(\frac{\eta}{\xi},\frac{Q^2}{\mu^2};\alpha_s(\mu)\right).
\end{align}
\end{subequations}
Thus the full singlet vector CFF may also be written directly in terms of the input quark and
gluon conformal moments at $\mu_0$ as
\begin{equation}
\mathcal H_{\rm pQCD}^{\rm sing}(\xi,\eta,t,Q^2)
=
\sum_{j=1}^{\infty}
[1+(-1)^j]\,
\xi^{-j}
\left[
C_{j,\rm eff}^{\Sigma}\,H_j^\Sigma(\eta,t;\mu_0)
+
C_{j,\rm eff}^{g}\,H_j^g(\eta,t;\mu_0)
\right].
\label{eq:app-pqcd-H-Sigma-g-inputscale}
\end{equation}
This form makes the relation between the diagonal $\pm$ representation and the physical
quark--gluon basis completely explicit.

\subsection{Fixing the normalization constants \texorpdfstring{$c_j^\pm$}{cj pm} by matching to the forward amplitude}
\label{app:pqcd-fix-cpm}

The conformal-symmetry prediction fixes the \emph{functional} dependence of the nonforward
Wilson coefficients on \(\eta/\xi\), but it does not by itself fix the overall normalization
constants \(c_j^\pm\).  These constants are fixed by requiring that the nonforward coefficient
functions reduce, in the forward limit \(\eta\to0\), to the standard forward DIS Wilson
coefficients.

Indeed, since
\begin{equation}
{}_2F_1\!\left(
\frac{j}{2}+\frac{\gamma_j^\pm}{4},
\frac{j+1}{2}+\frac{\gamma_j^\pm}{4};
j+\frac32+\frac{\gamma_j^\pm}{2};
\frac{\eta^2}{\xi^2}
\right)
\;\xrightarrow[\eta\to0]{}\;1,
\end{equation}
the nonforward coefficients reduce to
\begin{equation}
C_j^\pm\!\left(0,\frac{Q^2}{\mu^2};\alpha_s\right)
=
c_j^\pm(\alpha_s)\times
\Bigl[\text{the corresponding scale factor}\Bigr].
\end{equation}
Hence the constants \(c_j^\pm\) are precisely the forward Wilson-coefficient normalizations
written in the \(\pm\) basis.

Let
\begin{equation}
\bm c_j^{(\Sigma g)}(\alpha_s)
\equiv
\bigl(c_j^\Sigma(\alpha_s),\,c_j^g(\alpha_s)\bigr)
\end{equation}
denote the forward Wilson-coefficient row vector in the singlet quark--gluon basis.  Since the
Wilson coefficients transform as row vectors, while the conformal moments transform as column
vectors, the basis change is
\begin{equation}
\bm c_j^{(\Sigma g)}(\alpha_s)
=
\bm c_j^{(\pm)}(\alpha_s)\,U_j,
\qquad
\bm c_j^{(\pm)}(\alpha_s)
\equiv
\bigl(c_j^+(\alpha_s),\,c_j^-(\alpha_s)\bigr),
\label{eq:app-pqcd-cpm-row-transform}
\end{equation}
with \(U_j\) given in \eqref{eq:app-pqcd-U-def}.  Therefore
\begin{equation}
\bm c_j^{(\pm)}(\alpha_s)
=
\bm c_j^{(\Sigma g)}(\alpha_s)\,U_j^{-1}.
\label{eq:app-pqcd-cpm-from-forward}
\end{equation}

Using
\begin{equation}
U_j^{-1}
=
\frac{1}{D_j}
\begin{pmatrix}
1 & -a_j\\[1ex]
-b_j & 1
\end{pmatrix},
\qquad
D_j=1-a_j b_j,
\end{equation}
one obtains the general matching formulas
\begin{subequations}
\label{eq:app-pqcd-cpm-general}
\begin{align}
c_j^+(\alpha_s)
&=
\frac{c_j^\Sigma(\alpha_s)-b_j\,c_j^g(\alpha_s)}{D_j},
\\[1ex]
c_j^-(\alpha_s)
&=
\frac{c_j^g(\alpha_s)-a_j\,c_j^\Sigma(\alpha_s)}{D_j}.
\end{align}
\end{subequations}
These equations are completely general in the present normalization convention.  In particular,
they show that once the forward DIS coefficients are known in the quark--gluon basis, the
constants \(c_j^\pm\) are fixed uniquely.

At LO, the virtual photon couples directly only to the singlet quark operator, so the forward
quark--gluon coefficient row vector is simply
\begin{equation}
\bm c_j^{(\Sigma g),(0)}=\bigl(c_j^{\Sigma,(0)},\,c_j^{g,(0)}\bigr)=(1,0).
\label{eq:app-pqcd-forward-LO-sigg}
\end{equation}
Therefore the LO normalization constants in the \(\pm\) basis are
\begin{subequations}
\label{eq:app-pqcd-cpm-LO}
\begin{align}
c_j^{+,(0)}
&=
\frac{1}{D_j},
\\[1ex]
c_j^{-,(0)}
&=
-\frac{a_j}{D_j}.
\end{align}
\end{subequations}
Using the definitions of \(a_j\), \(b_j\), and \(D_j\), these may be written in the equivalent forms
\begin{subequations}
\label{eq:app-pqcd-cpm-LO-explicit}
\begin{align}
c_j^{+,(0)}
&=
\frac{\gamma_j^{\Sigma\Sigma,(0)}-\gamma_j^{-,(0)}}{\gamma_j^{+,(0)}-\gamma_j^{-,(0)}}
=
\frac{\gamma_j^{+,(0)}-\gamma_j^{gg,(0)}}{\gamma_j^{+,(0)}-\gamma_j^{-,(0)}},
\\[1ex]
c_j^{-,(0)}
&=
-\frac{\gamma_j^{\Sigma g,(0)}}{\gamma_j^{+,(0)}-\gamma_j^{-,(0)}}.
\end{align}
\end{subequations}
Thus, within the normalization convention of \eqref{eq:app-pqcd-U-def}, the LO coefficients
\(c_j^\pm\) are \emph{not} equal to \(1\) and \(0\); rather, they are the entries of the forward
quark coefficient \((1,0)\) rewritten in the \(\pm\) basis.

Consequently, the strict tree-level nonforward Wilson coefficients are
\begin{equation}
C_j^{\pm,\mathrm{tree}}\!\left(\frac{\eta}{\xi}\right)
=
c_j^{\pm,(0)}
{}_2F_1\!\left(
\frac{j}{2},
\frac{j+1}{2};
j+\frac32;
\frac{\eta^2}{\xi^2}
\right).
\label{eq:app-pqcd-Cpm-tree}
\end{equation}
If one wishes instead to keep the LO anomalous-dimension improvement at a conformal fixed point
or with LO running coupling, the same normalization constants \(c_j^{\pm,(0)}\) multiply the
corresponding scale-dependent factors appearing in
\eqref{eq:app-pqcd-fixed-point-Wilson} and \eqref{eq:app-pqcd-LO-running-Wilson}.

Finally, we stress that these explicit values of \(c_j^\pm\) depend on the normalization convention
chosen for the \(\pm\) eigenvectors.  A different normalization of the eigenvectors rescales
\(c_j^\pm\) and \(H_j^\pm\) inversely, leaving the physical products \(C_j^\pm H_j^\pm\) and hence
the Compton form factor unchanged.

\subsection{Large--\texorpdfstring{$N_c$}{Nc} limit of the normalization constants \texorpdfstring{$c_j^\pm$}{cj pm}}
\label{app:pqcd-largeNc-cpm}

It is useful to make the large--$N_c$ behavior of the LO normalization constants
$c_j^{\pm,(0)}$ completely explicit.  Starting from
Eq.~\eqref{eq:app-pqcd-cpm-LO-explicit}, one may rewrite the LO coefficients as
\begin{subequations}
\label{eq:app-pqcd-cpm-LO-compact}
\begin{align}
c_j^{+,(0)}
&=
\frac12\left[
1+
\frac{\gamma_j^{\Sigma\Sigma,(0)}-\gamma_j^{gg,(0)}}
{\sqrt{\left(\gamma_j^{\Sigma\Sigma,(0)}-\gamma_j^{gg,(0)}\right)^2
+4\,\gamma_j^{\Sigma g,(0)}\gamma_j^{g\Sigma,(0)}}}
\right],
\\[1ex]
c_j^{-,(0)}
&=
-\frac{\gamma_j^{\Sigma g,(0)}}
{\sqrt{\left(\gamma_j^{\Sigma\Sigma,(0)}-\gamma_j^{gg,(0)}\right)^2
+4\,\gamma_j^{\Sigma g,(0)}\gamma_j^{g\Sigma,(0)}}}.
\end{align}
\end{subequations}

We now take the standard 't~Hooft large--$N_c$ limit,
\begin{equation}
N_c\to\infty,
\qquad
\lambda\equiv g^2 N_c\ \text{fixed},
\qquad
n_f\ \text{fixed}.
\end{equation}
Then
\begin{equation}
C_F=\frac{N_c^2-1}{2N_c}
=
\frac{N_c}{2}+\ord\!\left(\frac{1}{N_c}\right),
\qquad
C_A=N_c,
\qquad
\beta_0=\frac{11}{3}N_c-\frac{2}{3}n_f.
\end{equation}
Hence the LO anomalous dimensions scale as
\begin{subequations}
\label{eq:app-pqcd-LO-AD-largeNc}
\begin{align}
\gamma_j^{\Sigma\Sigma,(0)}
&=
N_c\,\widehat{\gamma}_j^{\,q}
+\ord\!\left(\frac{1}{N_c}\right),
\qquad
\widehat{\gamma}_j^{\,q}
=
2S_1(j)-\frac32-\frac{1}{j(j+1)},
\\[1ex]
\gamma_j^{gg,(0)}
&=
N_c\,\widehat{\gamma}_j^{\,g}
+\ord(N_c^0),
\qquad
\widehat{\gamma}_j^{\,g}
=
4S_1(j)-\frac{11}{3}
+\frac{4}{j(j+1)}
-\frac{12}{(j-1)(j+2)},
\\[1ex]
\gamma_j^{\Sigma g,(0)}
&=
-2n_f\,r_j+\ord\!\left(\frac{1}{N_c}\right),
\qquad
r_j\equiv \frac{j^2+j+2}{j(j+1)(j+2)},
\\[1ex]
\gamma_j^{g\Sigma,(0)}
&=
-N_c\,s_j+\ord\!\left(\frac{1}{N_c}\right),
\qquad
s_j\equiv \frac{j^2+j+2}{(j-1)j(j+1)}.
\end{align}
\end{subequations}
It is convenient to introduce
\begin{equation}
\Omega_j
\equiv
\widehat{\gamma}_j^{\,q}-\widehat{\gamma}_j^{\,g}
=
\frac{13}{6}
-2S_1(j)
-\frac{5}{j(j+1)}
+\frac{12}{(j-1)(j+2)},
\label{eq:app-pqcd-Omegaj}
\end{equation}
and
\begin{equation}
R_j\equiv r_j s_j
=
\frac{(j^2+j+2)^2}{(j-1)j^2(j+1)^2(j+2)}.
\label{eq:app-pqcd-Rj}
\end{equation}
Then the discriminant in \eqref{eq:app-pqcd-cpm-LO-compact} becomes
\begin{align}
&\left(\gamma_j^{\Sigma\Sigma,(0)}-\gamma_j^{gg,(0)}\right)^2
+4\,\gamma_j^{\Sigma g,(0)}\gamma_j^{g\Sigma,(0)}
\nonumber\\[1ex]
&=
N_c^2\,\Omega_j^2
+
8n_f N_c\,R_j
+\ord(N_c^0),
\end{align}
and therefore
\begin{equation}
\sqrt{
\left(\gamma_j^{\Sigma\Sigma,(0)}-\gamma_j^{gg,(0)}\right)^2
+4\,\gamma_j^{\Sigma g,(0)}\gamma_j^{g\Sigma,(0)}
}
=
N_c|\Omega_j|
\left[
1+\frac{4n_f R_j}{N_c\,\Omega_j^2}
+\ord\!\left(\frac{1}{N_c^2}\right)
\right].
\label{eq:app-pqcd-discriminant-largeNc}
\end{equation}

Substituting \eqref{eq:app-pqcd-discriminant-largeNc} into
\eqref{eq:app-pqcd-cpm-LO-compact} yields the large--$N_c$ expansion
\begin{subequations}
\label{eq:app-pqcd-cpm-largeNc-general}
\begin{align}
c_j^{+,(0)}
&=
\frac12\Bigl[1+\operatorname{sgn}(\Omega_j)\Bigr]
-\operatorname{sgn}(\Omega_j)\,
\frac{2n_f R_j}{N_c\,\Omega_j^2}
+\ord\!\left(\frac{1}{N_c^2}\right),
\\[1ex]
c_j^{-,(0)}
&=
\frac{2n_f}{N_c}\,
\frac{r_j}{|\Omega_j|}
+\ord\!\left(\frac{1}{N_c^2}\right).
\end{align}
\end{subequations}
Equivalently, in piecewise form,
\begin{subequations}
\label{eq:app-pqcd-cpm-largeNc-piecewise}
\begin{align}
\Omega_j>0:\qquad
c_j^{+,(0)}
&=
1-\frac{2n_f R_j}{N_c\,\Omega_j^2}
+\ord\!\left(\frac{1}{N_c^2}\right),
&
c_j^{-,(0)}
&=
\frac{2n_f}{N_c}\frac{r_j}{\Omega_j}
+\ord\!\left(\frac{1}{N_c^2}\right),
\\[1ex]
\Omega_j<0:\qquad
c_j^{+,(0)}
&=
\frac{2n_f R_j}{N_c\,\Omega_j^2}
+\ord\!\left(\frac{1}{N_c^2}\right),
&
c_j^{-,(0)}
&=
\frac{2n_f}{N_c}\frac{r_j}{|\Omega_j|}
+\ord\!\left(\frac{1}{N_c^2}\right).
\end{align}
\end{subequations}

For the even-$j$ vector sector relevant here, one finds
\begin{equation}
\Omega_2=\frac43>0,
\qquad
\Omega_j<0 \quad \text{for all even } j\ge 4.
\end{equation}
Thus the lowest nontrivial moment is special:
\begin{subequations}
\label{eq:app-pqcd-cpm-largeNc-j2}
\begin{align}
c_2^{+,(0)}
&=
1-\frac{n_f}{2N_c}
+\ord\!\left(\frac{1}{N_c^2}\right),
\\[1ex]
c_2^{-,(0)}
&=
\frac{n_f}{2N_c}
+\ord\!\left(\frac{1}{N_c^2}\right).
\end{align}
\end{subequations}
By contrast, for even \(j\ge4\) one has
\begin{subequations}
\label{eq:app-pqcd-cpm-largeNc-jge4}
\begin{align}
c_j^{+,(0)}
&=
\frac{2n_f R_j}{N_c\,\Omega_j^2}
+\ord\!\left(\frac{1}{N_c^2}\right),
\\[1ex]
c_j^{-,(0)}
&=
\frac{2n_f}{N_c}\,
\frac{r_j}{|\Omega_j|}
+\ord\!\left(\frac{1}{N_c^2}\right).
\end{align}
\end{subequations}

Finally, one should keep in mind that in the normalization convention adopted in
\eqref{eq:app-pqcd-U-def}, the labels \(+\) and \(-\) do not coincide uniformly with a purely
quark-like or gluon-like channel at large \(N_c\).  In particular, when \(\Omega_j<0\) the entry
\begin{equation}
b_j=
\frac{\gamma_j^{\Sigma\Sigma,(0)}-\gamma_j^{+,(0)}}{\gamma_j^{\Sigma g,(0)}}
\sim N_c
\end{equation}
is large, so the eigenmoment \(H_j^-\) carries an \(\ord(N_c)\) quark admixture.  Therefore the
physically relevant quantities are the products \(c_j^\pm H_j^\pm\), which remain finite and are
independent of the normalization convention chosen for the \(\pm\) eigenvectors.

\subsection{Expansion of \texorpdfstring{$\gamma_j^{\pm,(0)}$}{gamma pm} near \texorpdfstring{$j=2$}{j=2}}
\label{app:pqcd-gamma-pm-near-j2}

It is useful to have the LO singlet eigen-anomalous dimensions expanded analytically around the
first nontrivial even spin \(j=2\).  We therefore set
\begin{equation}
j=2+u,
\qquad
|u|\ll1,
\label{eq:app-pqcd-j-2-plus-u}
\end{equation}
and expand the harmonic sum as
\begin{equation}
S_1(2+u)
=
\frac32
+
\left(\frac{\pi^2}{6}-\frac54\right)u
+\ord(u^2).
\label{eq:app-pqcd-S1-near2}
\end{equation}

Using the LO singlet anomalous-dimension entries introduced above, one finds
\begin{subequations}
\label{eq:app-pqcd-LO-entries-near2}
\begin{align}
\gamma_{2+u}^{\Sigma\Sigma,(0)}
&=
\frac{8}{3}C_F
+
\frac{C_F}{18}\left(12\pi^2-85\right)u
+\ord(u^2),
\\[1ex]
\gamma_{2+u}^{\Sigma g,(0)}
&=
-\frac{4}{3}T_F n_f
+
\frac{11}{18}T_F n_f\,u
+\ord(u^2),
\\[1ex]
\gamma_{2+u}^{g\Sigma,(0)}
&=
-\frac{8}{3}C_F
+
\frac{29}{9}C_F\,u
+\ord(u^2),
\\[1ex]
\gamma_{2+u}^{gg,(0)}
&=
\frac{4}{3}T_F n_f
+
\frac{C_A}{36}\left(24\pi^2-65\right)u
+\ord(u^2).
\end{align}
\end{subequations}

It is convenient to introduce the trace and determinant of the LO singlet matrix,
\begin{equation}
T_j^{(0)}
\equiv
\gamma_j^{\Sigma\Sigma,(0)}+\gamma_j^{gg,(0)},
\qquad
D_j^{(0)}
\equiv
\gamma_j^{\Sigma\Sigma,(0)}\gamma_j^{gg,(0)}
-
\gamma_j^{\Sigma g,(0)}\gamma_j^{g\Sigma,(0)}.
\end{equation}
Near \(j=2\), these become
\begin{subequations}
\label{eq:app-pqcd-trace-det-near2}
\begin{align}
T_{2+u}^{(0)}
&=
T_0+T_1\,u+\ord(u^2),
\\
D_{2+u}^{(0)}
&=
D_1\,u+\ord(u^2),
\end{align}
\end{subequations}
with
\begin{subequations}
\label{eq:app-pqcd-T0T1D1}
\begin{align}
T_0
&=
\frac{4}{3}\left(2C_F+T_F n_f\right),
\\[1ex]
T_1
&=
\frac{C_F}{18}\left(12\pi^2-85\right)
+
\frac{C_A}{36}\left(24\pi^2-65\right),
\\[1ex]
D_1
&=
\frac{2C_F}{27}
\left[
C_A\left(24\pi^2-65\right)
+
T_F n_f\left(12\pi^2-5\right)
\right].
\end{align}
\end{subequations}

Since the LO eigenvalues satisfy
\begin{equation}
\gamma_j^{+,(0)}+\gamma_j^{-,(0)}=T_j^{(0)},
\qquad
\gamma_j^{+,(0)}\gamma_j^{-,(0)}=D_j^{(0)},
\label{eq:app-pqcd-sum-prod-eigenvalues}
\end{equation}
and since \(D_2^{(0)}=0\), one obtains immediately
\begin{equation}
\gamma_2^{-,(0)}=0,
\qquad
\gamma_2^{+,(0)}=T_0=\frac{4}{3}\left(2C_F+T_F n_f\right).
\label{eq:app-pqcd-gamma-pm-at2}
\end{equation}
The vanishing of \(\gamma_2^{-,(0)}\) is the LO manifestation of the singlet momentum-sum rule.

To first order in \(u=j-2\), the eigenvalues therefore expand as
\begin{subequations}
\label{eq:app-pqcd-gamma-pm-near2}
\begin{align}
\gamma_{2+u}^{-,(0)}
&=
\frac{D_1}{T_0}\,u+\ord(u^2)
\nonumber\\
&=
\frac{C_F\Bigl[
C_A\left(24\pi^2-65\right)
+
T_F n_f\left(12\pi^2-5\right)
\Bigr]}
{18\left(2C_F+T_F n_f\right)}
\,(j-2)
+\ord\!\big((j-2)^2\big),
\\[2ex]
\gamma_{2+u}^{+,(0)}
&=
T_0+\left(T_1-\frac{D_1}{T_0}\right)u+\ord(u^2)
\nonumber\\
&=
\frac{4}{3}\left(2C_F+T_F n_f\right)
+
\frac{
4C_F^2\left(12\pi^2-85\right)
+
T_F n_f\Bigl[
C_A\left(24\pi^2-65\right)-160C_F
\Bigr]
}
{36\left(2C_F+T_F n_f\right)}
\,(j-2)
+\ord\!\big((j-2)^2\big).
\end{align}
\end{subequations}

Thus, near \(j=2\) the minus eigenvalue starts linearly from zero, while the plus eigenvalue
remains finite and receives a regular linear correction.  These formulas are useful when analyzing
both the forward normalization constants \(c_j^\pm\) and the behavior of the singlet fixed--\(j\)
Compton amplitude in the vicinity of the first even spin.

\end{widetext}

\bibliographystyle{apsrev4-2}
\bibliography{references-DDVCS}

@article{Ji:1996ek,
  author        = "Ji, Xiangdong",
  title         = "{Deeply Virtual Compton Scattering}",
  eprint        = "hep-ph/9609381",
  archivePrefix = "arXiv",
  primaryClass  = "hep-ph",
  doi           = "10.1103/PhysRevD.55.7114",
  journal       = "Phys. Rev. D",
  volume        = "55",
  pages         = "7114--7125",
  year          = "1997"
}

@article{Radyushkin:1996ru,
  author        = "Radyushkin, A. V.",
  title         = "{Scaling Limit of Deeply Virtual Compton Scattering}",
  eprint        = "hep-ph/9604317",
  archivePrefix = "arXiv",
  primaryClass  = "hep-ph",
  doi           = "10.1016/0370-2693(96)00528-X",
  journal       = "Phys. Lett. B",
  volume        = "380",
  pages         = "417--425",
  year          = "1996"
}

@article{Radyushkin:1996nd,
  author        = "Radyushkin, A. V.",
  title         = "{Asymmetric Gluon Distributions and Hard Diffractive Electroproduction}",
  eprint        = "hep-ph/9605431",
  archivePrefix = "arXiv",
  primaryClass  = "hep-ph",
  doi           = "10.1016/0370-2693(96)00844-1",
  journal       = "Phys. Lett. B",
  volume        = "385",
  pages         = "333--342",
  year          = "1996"
}

@article{Radyushkin:1997ki,
  author        = "Radyushkin, A. V.",
  title         = "{Nonforward Parton Distributions}",
  eprint        = "hep-ph/9704207",
  archivePrefix = "arXiv",
  primaryClass  = "hep-ph",
  journal       = "Phys. Rev. D",
  volume        = "56",
  pages         = "5524--5557",
  year          = "1997",
  doi           = "10.1103/PhysRevD.56.5524"
}

@article{Ji:1998xh,
  author        = "Ji, Xiangdong and Osborne, Jonathan",
  title         = "{One-Loop Corrections and All Order Factorization In Deeply Virtual Compton Scattering}",
  eprint        = "hep-ph/9801260",
  archivePrefix = "arXiv",
  primaryClass  = "hep-ph",
  doi           = "10.1103/PhysRevD.58.094018",
  journal       = "Phys. Rev. D",
  volume        = "58",
  pages         = "094018",
  year          = "1998"
}

@article{Collins:1998be,
  author        = "Collins, John C. and Freund, Andreas",
  title         = "{Proof of Factorization for Deeply Virtual Compton Scattering in QCD}",
  eprint        = "hep-ph/9801262",
  archivePrefix = "arXiv",
  primaryClass  = "hep-ph",
  doi           = "10.1103/PhysRevD.59.074009",
  journal       = "Phys. Rev. D",
  volume        = "59",
  pages         = "074009",
  year          = "1999"
}

@article{Diehl:2003ny,
  author        = "Diehl, Markus",
  title         = "{Generalized Parton Distributions}",
  eprint        = "hep-ph/0307382",
  archivePrefix = "arXiv",
  primaryClass  = "hep-ph",
  doi           = "10.1016/j.physrep.2003.08.002",
  journal       = "Phys. Rept.",
  volume        = "388",
  pages         = "41--277",
  year          = "2003"
}

@article{Belitsky:2005qn,
  author        = "Belitsky, A. V. and Radyushkin, A. V.",
  title         = "{Unraveling Hadronic Structure with Generalized Parton Distributions}",
  eprint        = "hep-ph/0504030",
  archivePrefix = "arXiv",
  primaryClass  = "hep-ph",
  doi           = "10.1016/j.physrep.2005.06.002",
  journal       = "Phys. Rept.",
  volume        = "418",
  pages         = "1--387",
  year          = "2005"
}

@article{Mueller:1997hs,
  author        = "M{\"u}ller, Dieter",
  title         = "{Restricted Conformal Invariance in QCD and Its Predictive Power for Virtual Two-Photon Processes}",
  eprint        = "hep-ph/9704406",
  archivePrefix = "arXiv",
  primaryClass  = "hep-ph",
  doi           = "10.1103/PhysRevD.58.054005",
  journal       = "Phys. Rev. D",
  volume        = "58",
  pages         = "054005",
  year          = "1998"
}

@article{Belitsky:1997rh,
  author        = "Belitsky, A. V. and M{\"u}ller, Dieter",
  title         = "{Predictions from Conformal Algebra for the Deeply Virtual Compton Scattering}",
  eprint        = "hep-ph/9709379",
  archivePrefix = "arXiv",
  primaryClass  = "hep-ph",
  doi           = "10.1016/S0370-2693(97)01390-7",
  journal       = "Phys. Lett. B",
  volume        = "417",
  pages         = "129--140",
  year          = "1998"
}

@article{Mueller:2005ed,
  author        = "M{\"u}ller, Dieter and Sch{\"a}fer, Andreas",
  title         = "{Complex Conformal Spin Partial Wave Expansion of Generalized Parton Distributions and Distribution Amplitudes}",
  eprint        = "hep-ph/0509204",
  archivePrefix = "arXiv",
  primaryClass  = "hep-ph",
  doi           = "10.1016/j.nuclphysb.2006.01.019",
  journal       = "Nucl. Phys. B",
  volume        = "739",
  pages         = "1--59",
  year          = "2006"
}

@article{Kirch:2005hu,
  author        = "Kirch, M. and Manashov, A. and Sch{\"a}fer, A.",
  title         = "{Evolution Equation for Generalized Parton Distributions}",
  eprint        = "hep-ph/0509330",
  archivePrefix = "arXiv",
  primaryClass  = "hep-ph",
  doi           = "10.1103/PhysRevD.72.114006",
  journal       = "Phys. Rev. D",
  volume        = "72",
  pages         = "114006",
  year          = "2005"
}

@article{Manashov:2005qm,
  author        = "Manashov, A. and Kirch, M. and Sch{\"a}fer, A.",
  title         = "{Solving the Leading Order Evolution Equation for GPDs}",
  eprint        = "hep-ph/0503109",
  archivePrefix = "arXiv",
  primaryClass  = "hep-ph",
  doi           = "10.1103/PhysRevLett.95.012002",
  journal       = "Phys. Rev. Lett.",
  volume        = "95",
  pages         = "012002",
  year          = "2005"
}

@article{Kumericki:2006xx,
  author        = "Kumeri\v{c}ki, K. and M{\"u}ller, D. and Passek-Kumeri\v{c}ki, K. and Sch{\"a}fer, A.",
  title         = "{Deeply Virtual Compton Scattering Beyond Next-to-Leading Order: The Flavor Singlet Case}",
  eprint        = "hep-ph/0605237",
  archivePrefix = "arXiv",
  primaryClass  = "hep-ph",
  doi           = "10.1016/j.physletb.2007.02.071",
  journal       = "Phys. Lett. B",
  volume        = "648",
  pages         = "186--194",
  year          = "2007"
}

@article{Kumericki:2007sa,
  author        = "Kumeri\v{c}ki, K. and M{\"u}ller, D. and Passek-Kumeri\v{c}ki, K.",
  title         = "{Towards a Fitting Procedure for Deeply Virtual Compton Scattering at Next-to-Leading Order and Beyond}",
  eprint        = "hep-ph/0703179",
  archivePrefix = "arXiv",
  primaryClass  = "hep-ph",
  doi           = "10.1016/j.nuclphysb.2007.10.029",
  journal       = "Nucl. Phys. B",
  volume        = "794",
  pages         = "244--323",
  year          = "2008"
}

@article{Braun:2020yib,
  author        = "Braun, V. M. and Manashov, A. N. and Moch, S. and Schoenleber, J.",
  title         = "{Two-Loop Coefficient Function for DVCS: Vector Contributions}",
  eprint        = "2007.06348",
  archivePrefix = "arXiv",
  primaryClass  = "hep-ph",
  doi           = "10.1007/JHEP09(2020)117",
  journal       = "JHEP",
  volume        = "09",
  pages         = "117",
  year          = "2020",
  note          = "Erratum: JHEP 02 (2022) 115"
}

@article{Braun:2021ysj,
  author        = "Braun, V. M. and Manashov, A. N. and Moch, S. and Schoenleber, J.",
  title         = "{The Axial-Vector Contributions in Two-Photon Reactions: Pion Transition Form Factor and Deeply-Virtual Compton Scattering at NNLO in QCD}",
  eprint        = "2106.01437",
  archivePrefix = "arXiv",
  primaryClass  = "hep-ph",
  doi           = "10.1103/PhysRevD.104.094007",
  journal       = "Phys. Rev. D",
  volume        = "104",
  pages         = "094007",
  year          = "2021"
}

@article{Braun:2022nnlo,
  author        = "Braun, V. M. and Ji, Yao and Schoenleber, Jakob",
  title         = "{Deeply-Virtual Compton Scattering at the Next-to-Next-to-Leading Order}",
  eprint        = "2207.06818",
  archivePrefix = "arXiv",
  primaryClass  = "hep-ph",
  doi           = "10.1103/PhysRevLett.129.172001",
  journal       = "Phys. Rev. Lett.",
  volume        = "129",
  pages         = "172001",
  year          = "2022"
}

@article{Ji:2023transversity,
  author        = "Ji, Yao and Schoenleber, Jakob",
  title         = "{Two-Loop Coefficient Functions in Deeply Virtual Compton Scattering: Flavor-Singlet Axial-Vector and Transversity Case}",
  eprint        = "2310.05724",
  archivePrefix = "arXiv",
  primaryClass  = "hep-ph",
  doi           = "10.1007/JHEP01(2024)053",
  journal       = "JHEP",
  volume        = "01",
  pages         = "053",
  year          = "2024"
}

@article{Braun:2024ddvcs,
  author        = "Braun, Vladimir M. and Jiang, Hua-Yu and Manashov, Alexander N. and von Manteuffel, Andreas",
  title         = "{The Two-Loop Coefficient Functions for Double Deeply Virtual Compton Scattering}",
  eprint        = "2411.14985",
  archivePrefix = "arXiv",
  primaryClass  = "hep-ph",
  doi           = "10.1007/JHEP01(2025)069",
  journal       = "JHEP",
  volume        = "01",
  pages         = "069",
  year          = "2025"
}

@article{Braun:2026dvcsmom,
  author        = "Braun, Vladimir M. and Gotzler, Patrick and Manashov, Alexander N.",
  title         = "{Conformal Moments of the Two-Loop Coefficient Functions in DVCS}",
  eprint        = "2512.14295",
  archivePrefix = "arXiv",
  primaryClass  = "hep-ph",
  doi           = "10.1103/PhysRevD.113.074005",
  journal       = "Phys. Rev. D",
  volume        = "113",
  pages         = "074005",
  year          = "2026"
}

@article{Erlich:2005qh,
  author        = "Erlich, Joshua and Katz, Emanuel and Son, Dam T. and Stephanov, Mikhail A.",
  title         = "{QCD and a Holographic Model of Hadrons}",
  eprint        = "hep-ph/0501128",
  archivePrefix = "arXiv",
  primaryClass  = "hep-ph",
  doi           = "10.1103/PhysRevLett.95.261602",
  journal       = "Phys. Rev. Lett.",
  volume        = "95",
  pages         = "261602",
  year          = "2005"
}

@article{Grigoryan:2007vg,
  author        = "Grigoryan, H. R. and Radyushkin, A. V.",
  title         = "{Structure of Vector Mesons in a Holographic Model with Linear Confinement}",
  eprint        = "0706.1543",
  archivePrefix = "arXiv",
  primaryClass  = "hep-ph",
  doi           = "10.1103/PhysRevD.76.095007",
  journal       = "Phys. Rev. D",
  volume        = "76",
  pages         = "095007",
  year          = "2007"
}

@article{Brower:2006ea,
  author        = "Brower, Richard C. and Polchinski, Joseph and Strassler, Matthew J. and Tan, Chung-I",
  title         = "{The Pomeron and Gauge/String Duality}",
  eprint        = "hep-th/0603115",
  archivePrefix = "arXiv",
  primaryClass  = "hep-th",
  doi           = "10.1088/1126-6708/2007/12/005",
  journal       = "JHEP",
  volume        = "12",
  pages         = "005",
  year          = "2007"
}

@article{Costa:2012cb,
  author        = "Costa, Miguel S. and Djuri\'c, Marko",
  title         = "{Deeply Virtual Compton Scattering from Gauge/Gravity Duality}",
  eprint        = "1201.1307",
  archivePrefix = "arXiv",
  primaryClass  = "hep-th",
  doi           = "10.1103/PhysRevD.86.016009",
  journal       = "Phys. Rev. D",
  volume        = "86",
  pages         = "016009",
  year          = "2012"
}

@article{Brower:2012mk,
  author        = "Brower, Richard C. and Costa, Miguel S. and Djuri\'c, Marko and Tan, Chung-I",
  title         = "{Deeply Virtual Compton Scattering and Higgs Production Using the Pomeron in AdS}",
  eprint        = "1210.1511",
  archivePrefix = "arXiv",
  primaryClass  = "hep-ph",
  journal       = "PoS",
  volume        = "QCD-TNT-II",
  pages         = "028",
  year          = "2011"
}

@article{Nishio:2014rya,
    author = "Nishio, Ryoichi and Watari, Taizan",
    title = "{Skewness dependence of generalized parton distributions, conformal OPE, and the AdS/CFT correspondence}",
    doi = "10.1103/PhysRevD.90.125001",
    journal = "Phys. Rev. D",
    volume = "90",
    number = "12",
    pages = "125001",
    year = "2014"
}

@article{Hatta:2018jpsi,
  author        = "Hatta, Yoshitaka and Yang, Di-Lun",
  title         = "{Holographic $J/\psi$ Production Near-Threshold and the Proton Mass Problem}",
  eprint        = "1808.02163",
  archivePrefix = "arXiv",
  primaryClass  = "hep-ph",
  doi           = "10.1103/PhysRevD.98.074003",
  journal       = "Phys. Rev. D",
  volume        = "98",
  number        = "7",
  pages         = "074003",
  year          = "2018"
}

@article{Mamo:2019mka,
  author        = "Mamo, Kiminad A. and Zahed, Ismail",
  title         = "{Diffractive Photoproduction of $J/\psi$ and $\\Upsilon$ Using Holographic QCD: Gravitational Form Factors and GPD of Gluons in the Proton}",
  eprint        = "1910.04707",
  archivePrefix = "arXiv",
  primaryClass  = "hep-ph",
  doi           = "10.1103/PhysRevD.101.086003",
  journal       = "Phys. Rev. D",
  volume        = "101",
  number        = "8",
  pages         = "086003",
  year          = "2020"
}

@article{Guo:2022ump0,
  author        = "Guo, Yuxun and Ji, Xiangdong and Shiells, Kyle",
  title         = "{Generalized Parton Distributions through Universal Moment Parameterization: Zero Skewness Case}",
  eprint        = "2207.05768",
  archivePrefix = "arXiv",
  primaryClass  = "hep-ph",
  doi           = "10.1007/JHEP09(2022)215",
  journal       = "JHEP",
  volume        = "09",
  pages         = "215",
  year          = "2022"
}

@article{Guo:2023ump,
  author        = "Guo, Yuxun and Ji, Xiangdong and Santiago, M. Gabriel and Shiells, Kyle and Yang, Jinghong",
  title         = "{Generalized Parton Distributions through Universal Moment Parameterization: Non-Zero Skewness Case}",
  eprint        = "2302.07279",
  archivePrefix = "arXiv",
  primaryClass  = "hep-ph",
  doi           = "10.1007/JHEP05(2023)150",
  journal       = "JHEP",
  volume        = "05",
  pages         = "150",
  year          = "2023"
}

@article{Guo:2025muf,
    author = "Guo, Yuxun and Aslan, Fatma P. and Ji, Xiangdong and Santiago, M. Gabriel",
    title = "{First Global Extraction of Generalized Parton Distributions from Experiment and Lattice Data with Next-to-Leading-Order Accuracy}",
    eprint = "2509.08037",
    archivePrefix = "arXiv",
    primaryClass = "hep-ph",
    doi = "10.1103/qct5-y7rp",
    journal = "Phys. Rev. Lett.",
    volume = "135",
    number = "26",
    pages = "261903",
    year = "2025"
}

@article{Mamo:2024vjh,
    author = "Mamo, Kiminad A. and Zahed, Ismail",
    title = "{String-based parametrization of nucleon GPDs at any skewness: A comparison to lattice QCD}",
    eprint = "2404.13245",
    archivePrefix = "arXiv",
    primaryClass = "hep-ph",
    doi = "10.1103/PhysRevD.110.114016",
    journal = "Phys. Rev. D",
    volume = "110",
    number = "11",
    pages = "114016",
    year = "2024"
}

@article{Mamo:2024jwp,
    author = "Mamo, Kiminad A. and Zahed, Ismail",
    title = "{Parametrization of Generalized Parton Distributions from t-Channel String Exchange in AdS Spaces}",
    eprint = "2411.04162",
    archivePrefix = "arXiv",
    primaryClass = "hep-ph",
    doi = "10.1103/PhysRevLett.133.241901",
    journal = "Phys. Rev. Lett.",
    volume = "133",
    number = "24",
    pages = "241901",
    year = "2024"
}

@article{Hechenberger:2025rye,
    author = "Hechenberger, Florian and Mamo, Kiminad A. and Zahed, Ismail",
    title = "{Rapidity-dependent spin decomposition of the nucleon}",
    eprint = "2507.18615",
    archivePrefix = "arXiv",
    primaryClass = "hep-ph",
    doi = "10.1103/4rqm-dxz2",
    journal = "Phys. Rev. D",
    volume = "113",
    number = "3",
    pages = "034027",
    year = "2026"
}

@article{Hechenberger:2025wnz,
    author = "Hechenberger, Florian and Mamo, Kiminad A. and Zahed, Ismail",
    title = "{String-based axial and helicity-flip GPDs: A comparison to lattice QCD}",
    eprint = "2508.00817",
    archivePrefix = "arXiv",
    primaryClass = "hep-ph",
    doi = "10.1103/7lxw-g6tq",
    journal = "Phys. Rev. D",
    volume = "112",
    number = "7",
    pages = "074018",
    year = "2025"
}

@article{Katz:2005ir,
    author = "Katz, Emanuel and Lewandowski, Adam and Schwartz, Matthew D.",
    title = "{Tensor mesons in AdS/QCD}",
    eprint = "hep-ph/0510388",
    archivePrefix = "arXiv",
    reportNumber = "UCB-PTH-05-37, LBNL-59061, BUHEP-05-16",
    doi = "10.1103/PhysRevD.74.086004",
    journal = "Phys. Rev. D",
    volume = "74",
    pages = "086004",
    year = "2006"
}

@article{Mager:2025pvz,
    author = "Mager, Jonas and Cappiello, Luigi and Leutgeb, Josef and Rebhan, Anton",
    title = "{Longitudinal Short-Distance Constraints on Hadronic Light-by-Light Scattering and Tensor-Meson Contributions to the Muon g-2}",
    eprint = "2501.19293",
    archivePrefix = "arXiv",
    primaryClass = "hep-ph",
    doi = "10.1103/dxwr-gpsl",
    journal = "Phys. Rev. Lett.",
    volume = "135",
    number = "9",
    pages = "091901",
    year = "2025"
}

@article{Ji:1996ek2,
    author = "Ji, Xiang-Dong",
    title = "{Gauge-Invariant Decomposition of Nucleon Spin}",
    eprint = "hep-ph/9603249",
    archivePrefix = "arXiv",
    reportNumber = "MIT-CTP-2517",
    doi = "10.1103/PhysRevLett.78.610",
    journal = "Phys. Rev. Lett.",
    volume = "78",
    pages = "610--613",
    year = "1997"
}

@misc{Mamo:2026vuq,
  author        = {Mamo, Kiminad A.},
  title         = {From Vacuum to Nucleon: Fixed--$j$ Kernel Matching of Holographic Current Correlators to QCD},
  year          = {2026},
  eprint        = {2604.12037},
  archivePrefix = {arXiv},
  primaryClass  = {hep-th}
}

@misc{Nishio:2014eua,
    author = "Nishio, Ryoichi and Watari, Taizan",
    title = "{Skewness Dependence of GPD / DVCS, Conformal OPE and AdS/CFT Correspondence II: a holographic model of GPD}",
    eprint = "1408.6365",
    archivePrefix = "arXiv",
    primaryClass = "hep-ph",
    reportNumber = "UT-14-38, IPMU14-291",
    month = "8",
    year = "2014"
}

@article{Bertone:2021yyz,
    author = "Bertone, V. and Dutrieux, H. and Mezrag, C. and Moutarde, H. and Sznajder, P.",
    title = "{Deconvolution problem of deeply virtual Compton scattering}",
    eprint = "2104.03836",
    archivePrefix = "arXiv",
    primaryClass = "hep-ph",
    doi = "10.1103/PhysRevD.103.114019",
    journal = "Phys. Rev. D",
    volume = "103",
    number = "11",
    pages = "114019",
    year = "2021"
}

@article{Marquet:2010sf,
    author = "Marquet, Cyrille and Roiesnel, Claude and Wallon, Samuel",
    title = "{Virtual Compton Scattering off a Spinless Target in AdS/QCD}",
    eprint = "1002.0566",
    archivePrefix = "arXiv",
    primaryClass = "hep-ph",
    doi = "10.1007/JHEP04(2010)051",
    journal = "JHEP",
    volume = "04",
    pages = "051",
    year = "2010"
}

@article{Deja:2023ahc,
    author = "Deja, K. and Martinez-Fernandez, V. and Pire, B. and Sznajder, P. and Wagner, J.",
    title = "{Phenomenology of double deeply virtual Compton scattering in the era of new experiments}",
    eprint = "2303.13668",
    archivePrefix = "arXiv",
    primaryClass = "hep-ph",
    doi = "10.1103/PhysRevD.107.094035",
    journal = "Phys. Rev. D",
    volume = "107",
    number = "9",
    pages = "094035",
    year = "2023"
}

@article{Alvarado:2025prc,
    author = "Alvarado, J. S. and Hoballah, M. and Voutier, E.",
    title = "{Sensitivity of double deeply virtual Compton-scattering observables to generalized parton distributions}",
    eprint = "2502.02346",
    archivePrefix = "arXiv",
    primaryClass = "hep-ph",
    doi = "10.1103/PhysRevC.111.065205",
    journal = "Phys. Rev. C",
    volume = "111",
    number = "6",
    pages = "065205",
    year = "2025"
}

@misc{Alvarado:2026ggy,
    author = "Alvarado, J. S. and others",
    title = "{Electro- and photoproduction of muon pairs with $\mu$CLAS12: Double Deeply Virtual Compton Scattering, Timelike Compton Scattering, and $J/\psi$ production}",
    eprint = "2605.11690",
    archivePrefix = "arXiv",
    primaryClass = "hep-ex",
    reportNumber = "JLAB-PHY-26-4699",
    month = "5",
    year = "2026"
}

@article{Accardi:2023chb,
    author = "Accardi, A. and others",
    title = "{Strong interaction physics at the luminosity frontier with 22 GeV electrons at Jefferson Lab}",
    eprint = "2306.09360",
    archivePrefix = "arXiv",
    primaryClass = "nucl-ex",
    reportNumber = "JLAB-PHY-23-3840, JLAB-THY-23-3848",
    doi = "10.1140/epja/s10050-024-01282-x",
    journal = "Eur. Phys. J. A",
    volume = "60",
    number = "9",
    pages = "173",
    year = "2024"
}

@article{Mamo:2021tzd,
    author = "Mamo, Kiminad A. and Zahed, Ismail",
    title = "{Electroproduction of heavy vector mesons using holographic QCD: From near threshold to high energy regimes}",
    eprint = "2106.00722",
    archivePrefix = "arXiv",
    primaryClass = "hep-ph",
    doi = "10.1103/PhysRevD.104.066023",
    journal = "Phys. Rev. D",
    volume = "104",
    number = "6",
    pages = "066023",
    year = "2021"
}

@article{Muller:1994ses,
    author = {M{\"u}ller, Dieter and Robaschik, D. and Geyer, B. and Dittes, F. -M. and Ho{\v{r}}ej{\v{s}}i, J.},
    title = "{Wave functions, evolution equations and evolution kernels from light ray operators of QCD}",
    eprint = "hep-ph/9812448",
    archivePrefix = "arXiv",
    reportNumber = "NTZ-6-91, NTZ-91-6",
    doi = "10.1002/prop.2190420202",
    journal = "Fortsch. Phys.",
    volume = "42",
    pages = "101--141",
    year = "1994"
}

\end{document}